\documentclass{aa}
\bibpunct{(}{)}{;}{a}{}{,} % to follow the A&A style
\usepackage{graphicx}
\usepackage{txfonts}
\usepackage{color}
\usepackage{booktabs}
\usepackage{float}
\usepackage{cancel}
\usepackage{amsmath}
\usepackage{soul}
\usepackage{lscape} 
\usepackage{longtable}
\usepackage[normalem]{ulem}
\bibpunct{(}{)}{;}{a}{}{,}
\usepackage{ae,aecompl}
\usepackage{color,soul}
\usepackage{adjustbox}
\usepackage{subfigure}
\usepackage{float}
\usepackage{tikz}
\usetikzlibrary{shapes,arrows}
\usepackage{natbib}
\usepackage{ulem}
\usepackage{comment}
\usepackage{amsmath}

\def\spitzer{{\it Spitzer\/}}

\def\acs606{\textit{F606W}}
\def\acs814{\textit{F814W}}

\def\irac36{${\rm 3.6\mu m}$}
\def\irac45{${\rm 4.5\mu m}$}
\def\irac58{${\rm 5.8\mu m}$}
\def\irac80{${\rm 8.0\mu m}$}
\def\mips24{${\rm 24\mu m}$}
\def\pep100{${\rm 100\mu m}$}
\def\pep160{${\rm 160\mu m}$}
\def\her250{${\rm 250\mu m}$}
\def\her350{${\rm 350\mu m}$}
\def\her500{${\rm 500\mu m}$}
\def\irac{{\rm IRAC\/}}
\def\mips{{\rm MIPS\/}}

\def\otelo{\hbox{OTELO}}
\def\chandra{\hbox{\textit{Chandra}}}
\def\acs{\hbox{HST-ACS}}

\def\pep{\hbox{PEP}}

\def\ha{H$\alpha$}
\def\hb{H$\beta$}

\def\oii{[\ion{O}{ii}]$\lambda\lambda$3726,3729} 
\def\oiiis{[\ion{O}{iii}]}
\def\oiii{[\ion{O}{iii}]$\lambda$4959,5007}

\def\oiis{[\ion{O}{ii}]}

\begin{document} 

\title{Star formation at z$\sim$0.9 from the OTELO survey:  A comprehensive view combining deep optical spectroscopy and infrared data}
\author{Rocío Navarro Martínez\inst{1,2,3},
Miguel Cerviño \inst{4},
Ricardo P\'erez-Martínez\inst{5,2},
Ana Mar\'ia P\'erez-Garc\'ia\inst{5,2},
Bernab\'e Cedr\'es\inst{6,7,2},
\'Angel Bongiovanni\inst{8,2},
Jakub Nadolny\inst{9,6,7},
Miguel S\'anchez-Portal \inst{8,2},
Jordi Cepa\inst{6,7,2},
Emilio Alfaro\inst{10},
Laia Barrufet\inst{14},
Jos\'e A. de Diego\inst{11},
Jes\'us Gallego\inst{3}, 
J. Jes\'us Gonz\'alez\inst{11},
Mauro Gonz\'alez-Otero\inst{10,2},
J. Ignacio Gonz\'alez-Serrano\inst{12,2},
and Carmen P. Padilla Torres\inst{6,7,13}}

\institute{\tiny
$^1$Escuela Superior de Ingeniería, Ciencia y Tecnología, UNIE Universidad, Arapiles 14 28015 Madrid, Spain\\
$^2$ Asociación Astrofísica para la Promoción de la Investigación, Instrumentación y su Desarrollo, ASPID, E-38205 La Laguna,
Tenerife, Spain \\
$^3$ Departamento de Física de la Tierra y Astrofísica, Instituto de Física de Partículas y del Cosmos, IPARCOS. Universidad
Complutense de Madrid, E-28040, Madrid, Spain\\
$^4$ Centro de Astrobiología (CAB), CSIC-INTA, Camino Bajo del Castillo s/n, 28692 Villanueva de la Cañada, Spain\\
$^5$ ISDEFE for European Space Astronomy Centre (ESAC)/ESA, P.O. Box 78, E-28690, Villanueva de la Cañada, Madrid, Spain\\
$^6$ Instituto de Astrofísica de Canarias (IAC), E-38200 La Laguna, Tenerife, Spain\\
$^7$ Departamento de Astrofísica, Universidad de La Laguna (ULL), E-38205 La Laguna, Tenerife, Spain\\
$^8$ Instituto de Radioastronomía Milimétrica (IRAM), Av. Divina Pastora 7, Núcleo Central, E-18012 Granada, Spain\\
$^9$ Astronomical Observatory Institute, Faculty of Physics and Astronomy, Adam Mickiewicz University, ul.~S{\l}oneczna 36, 60-286 Pozna{\'n}, Poland \\
$^{10}$ Instituto de Astrof\'isica de Andaluc\'ia, CSIC, E-18080, Granada, Spain\\
$^{11}$ Instituto de Astronom\'ia, Universidad Nacional Aut\'onoma de M\'exico, Apdo. Postal 70-264, 04510 Ciudad de M\'exico, Mexico\\
$^{12}$ Instituto de F\'isica de Cantabria (CSIC-Universidad de Cantabria), E-39005 Santander, Spain \\
$^{13}$Fundación Galileo Galilei - INAF Rambla José Ana Fernández Pérez, 7, 38712 Breña Baja,
Tenerife, Spain\\
$^14$ Institute for Astronomy, University of Edinburgh, Royal Observatory, Edinburgh EH9 3HJ, UK\\}

\date{Received ------ / Accepted -----}

% \abstract{}{}{}{}{} 
% 5 {} token are mandatory 
  \abstract
   {}
   {We aim to quantify the star formation rate (SFR) from deep optical spectroscopic data and far-infrared (FIR) photometry from a sample of galaxies at $z\sim 0.9$ from the \otelo\ survey and compare the activity estimated by optical tracers and FIR emission.   
   }
   {We used the multi-wavelength OTELO catalogue to construct a sample of FIR sources. We identified and separated galaxies with active nuclei and  
   derived the physical properties of the rest. We analysed their spectral energy distribution, obtaining estimates for stellar mass, dust attenuation, luminosity, and SFR based on infrared luminosity.
   We also studied \hb\ and \oiii\ emission-line galaxies without significant FIR emission from previous works.
   This approach allowed us to perform a comparative analysis among the SFR obtained through different calibrators, in particular \hb\, presented in a previous work}
   {We find that FIR-based SFR estimates uncover a significant fraction of hidden star formation. We determined that the SFR density obtained from the FIR emission is three times larger than that obtained from only emission-line sources. Likely related to the fact that each SFR tracer provides insight into star formation over different timescales, we suggest that such indicators are also more or less suited for different galaxy populations. Specifically, while optical emission lines effectively trace star formation in lower-mass galaxies, FIR-derived SFRs provide a more reliable measure in massive dust-rich systems. By accounting for both optically visible and obscured star formation, we provide a more comprehensive view of the star-forming main sequence at $z \sim 0.9$ and reinforce the importance of infrared tracers in studying galaxy evolution.}
   {}
   \titlerunning{Star formation at $z \sim 0.9$: Infrared and optical view}
   \authorrunning{R. Navarro et al.}
   \maketitle

\section{Introduction}
\label{sec:intro}

The estimation of the star formation rate (SFR), defined as the stellar mass formed per unit time, is a key parameter in the study of galaxy evolution. It provides critical insights into the processes governing the birth of stars and the growth of galaxies over cosmic time and directly relates to the mass assembly histories of galaxies, the enrichment of the interstellar medium, and the regulation of feedback processes that shape galaxy morphologies.

Traditionally, the SFR has been estimated using various calibrators, each covering different time ranges (\citeauthor{kennicutt2012star} \citeyear{kennicutt2012star}; see also \citeauthor{boquien2014impact} \citeyear{boquien2014impact} and \citeauthor{Ceretal16} \citeyear{Ceretal16} for a formal mathematical approach) and manifesting different strengths and limitations. One of the most used methods involves the study of emission lines to trace the presence of young massive stars. Lines such as \ha, \hb\,, or \oii\ provide a snapshot of the instantaneous SFR, as they are generated by the ionising radiation from O- and B-type stars \citep{kennicutt1998star}. However, they can be significantly affected by dust extinction and distort the estimate of the true star formation activity.

For galaxies that do not show clear emission lines in their spectra, alternative SFR estimators are required. Infrared emission serves as an ally in such scenarios. Dust absorbs ultraviolet and optical light from young stars and re-radiates it in IR wavelengths. This re-radiated IR emission provides a measure of the obscured SFR, avoiding the limitations raised by dust extinction that often affect optical observations \citep{Calzetti2000,kennicutt2009dust}. The IR emission therefore reveals hidden star formation activity that would otherwise remain undetected.

Works such as \cite{kewley2002halpha} have shown that the agreement between SFRs derived from IR and H$\alpha$ emission for both early- and late-type galaxies suggests that far-infrared (FIR) emission originates from the same young stellar populations responsible for the nebular emission lines. This interpretation has been further supported by physical correlations linking FIR emission to young stars, such as the radio-FIR correlation \citep{gavazzi1986radio} and the 12~$\mu$m–FIR luminosity correlation \citep{shapley2001multivariate}, implying that dust heating occurs in close proximity to star-forming regions. Additionally, the FIR–H$\alpha$ luminosity correlation points to a universal relationship between global dust and gas properties across all FIR-detected galaxies in the Nearby Field Galaxy Survey (NFGS). Together, these findings indicate that galaxies exhibiting line emission and those detected in the FIR are fundamentally similar in nature, differing primarily in the visibility of their star-forming regions rather than in intrinsic star formation efficiency. This provides a physical justification for comparing FIR-based and H$\beta$-derived SFRs in our study.

However, it is imperative that the IR emission accounted for does not have contamination from active galactic nuclei (AGNs) nor significant contribution from old stellar populations. We acknowledge that some AGNs host significant star formation activity; therefore, by excluding them, we may be underestimating the total SFR of the Universe. However, as we show in Sect. \ref{sec:Data}, the small AGN fraction found in our sample allows us to adopt this limitation, in agreement with previous works cited throughout this article.

Various galaxy characteristics have been obtained by spectral energy distribution (SED) fitting (see a review of this approach in \citealt{walcher2011fitting}). Among all the possible parameters that affect the galaxy spectra, dust is one of the most important since it can significantly alter the SED by absorbing and scattering stellar photons and emitting the absorbed energy in the IR range ($1 < \mathrm{\lambda} < 1000\, \mathrm{\mu m}$). This makes dust a crucial factor in the study of galaxy evolution.

Assuming an accurate extinction correction and using the Balmer decrement \citep{hirashita2003star}, H$\alpha$ luminosity is one of the most reliable SFR indicators. However, while optical surveys can miss a significant fraction of this obscured activity, the precise contribution of heavily dust-reddened galaxies is still being actively quantified. Moreover, there is evidence suggesting that the dust-obscured component has dominated the cosmic history of star formation for the past $\sim 12$ billion years, up to $z \sim 4$ \citep{zavala2021evolution}. Recent studies, particularly those based on deep \textit{Atacama Large Millimeter/submillimeter Array} (ALMA) and \textit{James Webb Space Telescope} (JWST) observations, have advanced our understanding of obscured star formation at various redshifts \citep{barrufet2023unveiling,traina2024a3cosmos, gentile2025going}. These recent works have suggested that dust-obscured galaxies play a pivotal role in galaxy evolution and may represent a substantial fraction of the star formation missed by optical surveys. In this context, properly accounting for the contribution of obscured galaxies remains essential to achieving a comprehensive understanding of the evolution of star formation across cosmic time.

In previous works \citep{Bongiovanni2020,NavarroMartinez2021}, we analysed the SFR in a sample of \hb\ and \oiii\ emission-line galaxies (ELGs) at $z\sim 0.9$ from the OSIRIS Tunable Emission Line Object (OTELO) survey.\footnote{{\tt http://research.iac.es/proyecto/otelo}}  In those studies, we defined ELGs as objects where (i) at least one point of the pseudo-spectrum lies above $f_c + 2\sigma_{med,c}$ and (ii) there is an adjacent point with a flux density above $f_c + \sigma_{med,c}$, where $f_c$ is the flux of the pseudo-continuum (the median flux of the pseudo-spectra) and $\sigma_{med,c}$ is the root of the averaged square deviation of the entire pseudospectrum with respect to $f_c$. We adopted the same criterion for the present work.

In this study, we focus on the obscured SFR in galaxies from the same survey that exhibit significant IR emission, considering both ELGs and galaxies without emission lines in their spectra. By using {\sc cigale} \citep{Boquien2019} for SED fitting, we aim to derive $M_\star$, dust attenuation, SFR, and IR luminosity ($L_{\mathrm{IR}}$) and to discuss the efficiency of $L_{\mathrm{IR}}$ emission as an SFR estimator in these sources.

Infrared-based SFR estimators have long been shown to reliably trace dust-obscured star formation, particularly in actively star-forming systems \citep{Elbaz2011,rodighiero2011lesser}, with Herschel observations demonstrating the tight link between $L_{\mathrm{IR}}$, the main sequence (MS) of star formation, and starburst outliers. Building on this legacy of Herschel-based studies, our investigation contributes to a broader understanding of galaxy evolution by shedding light on the obscured star formation activity of these otherwise elusive galaxies. By including IR observations and acknowledging possible AGN and old stellar population contributions to the IR emission, we aim to uncover the hidden star formation and provide a more complete picture of galaxy evolution.

This paper is structured as follows. In Sect. \ref{sec:Data} we describe the selection of FIR emitters in the OTELO survey and the AGN exclusion from this sample. Section \ref{sec:SEDfitting} addresses the fitting of the SED of the sample. In Sect. \ref{sec:Analysis} we present the physical parameters and star formation obtained from the previous fitting. A discussion of the results and a comparison with similar data are presented in Sect. \ref{sec:Discussion}. Finally, Sect. 6 reports the conclusions of this work.

Throughout this paper, we assume a standard $\Lambda$-cold dark matter cosmology with $\Omega_\Lambda = 0.7$, $\Omega_m = 0.3$, and $\mathrm{H}_0 = 70 \,\mathrm{km}\,  \mathrm{s}^{-1}\, \mathrm{Mpc}^{-1}$. All magnitudes are given in the AB system. We also assume a \cite{Chabrier2003} initial mass function (IMF), unless otherwise stated.

\section{Data selection}
\label{sec:Data}

For this study, we used data from the OTELO survey  
\cite[][hereafter OTELO-I]{Bongiovanni2019}  to extract a sample of FIR emitters at a redshift of approximately $z \sim 0.9$. We also took advantage of previous samples of \hb\ and \oiiis\ line emitting sources from the same survey analysed by \cite{NavarroMartinez2021} and \cite{Bongiovanni2020}, respectively, at a similar redshift. 

\subsection{The OTELO survey}
\label{otelosurvey}

The OTELO survey is a pencil beam survey employing the Red Tunable Filter (RTF) of the OSIRIS instrument \citep{Cepa03} of the 10.4m Gran Telescopio Canarias \cite[GTC;][]{Alvarez98}. 
This survey's design aims to address the limitations of slit-based spectroscopic surveys by sampling the spectral range of interest across its entire field of view, thereby generating pseudospectra with a resolution around $R \sim 700$, which is the flux-calibrated result of convolving the intrinsic SED of a source with the wavelength-dependent transmission response of the RTF scan, providing a low-resolution spectrum built by sampling successive narrowband slices.

Through the OTELO survey, a region with minimal sky emission lines of the Extended Groth Strip was observed in order to identify emission-line sources within various co-moving volumes, reaching redshifts up to 6.5. The RTF configuration involved scanning a window spanning $230~$\AA\ centred at $9175~$\AA\ within a $7.5 \times 7.4 ~\mathrm{arcmin}^2$ area situated at the south-western edge of Extended Groth Strip, a region extensively explored. The survey achieved a limiting line flux of 
$5 \times 10^{-19} \, \mathrm{erg}~\mathrm{s}^{-1}~\mathrm{cm}^{-2}$,
thus marking it as one of the deepest emission-line survey conducted to date. For a comprehensive overview of the survey, we refer to OTELO-I.

Utilising OSIRIS guaranteed time, a total of 108 hours was dedicated to acquiring OTELO data. From the co-added RTF images, we derived a source list, obtaining for each detected object a pseudo-spectrum composed of the 36 tomography slices that form the OTELO scan (see the criteria in \citealt{ramon2019otelo}). These measurements were enriched with publicly available data from the CFHTLS survey \cite[T0007 Relase;][]{cuillandre2012introduction}, \textit{Hubble} Space Telescope Advanced Camera for Surveys (HST-ACS), and near-infrared (NIR) data from the WIRcam Deep Survey \cite[WIRDS, Release T0002][]{bielby2012wircam} , together with mid-infrared (MIR) photometry from the \spitzer/\irac\ 3.6 and 4.5 $\mu \mathrm{m}$ bands \citep{Barro2011}, to establish the core catalogue. Additional datasets from X-ray from \chandra\ \citep{Weisskopf2002}, UV, MIR, and FIR catalogues were incorporated using refined cross-match techniques developed by \cite{PerezMartinez2016}, resulting in the final OTELO multi-wavelength catalogue with 11,237 raw entries.

Each of the 11,237 raw entries potentially provides an SED with up to 24 photometric datapoints. Those SEDs were used to derive photometric redshifts (photo-$z$'s) using {\sc LePhare} \citep{Arnouts1999, Ilbert2006}. This tool employs a galaxy template library that includes the four standard Hubble types \citep{Coleman1980} and six star-forming galaxy (SFG) templates \citep{Kinney1996}. This process is described in OTELO-I. The extinction law described by \cite{Calzetti2000} was applied, incorporating extinction values $\mathrm{E(B-V)}$ ranging from 0 to 1.1 in increments of 0.05. Both the value and quality of the photo-$z$ are incorporated into the final OTELO catalogue.
For more comprehensive insights into the construction of the final OTELO catalogue, extraction of pseudo-spectra, estimation of photometric redshifts, and the selection of emitting objects, readers are directed to OTELO-I and \cite{Bongiovanni2020}.

\subsection{Far-infrared sample selection}
\label{SampleSelection}
For the aim of this work, we have profited from the sample of OTELO emitters at \( 0.81 \le z \le 0.92 \). In this redshift range, the OTELO survey RTF wavelength coverage is tuned to observe redshfited \hb, and \oiiis\ emission. The \hb\ emitters from the sample are detailed in \cite{NavarroMartinez2021}, as well as \oiiis\ emitters in \cite{Bongiovanni2020}, respectively. In this spectral window \otelo\ may detect ELGs with \hb\ or \oiiis\ emission lines only. However there is an narrower redshift range, where both lines can appear in the PS. We identify 33 \hb-only, 170 \oiiis- only, and 12 \hb+\oiiis\ emission-line sources.

To construct the sample of FIR emitters with no emission line detected in OTELO, we first defined a redshift interval of \( 0.81 \le z_\mathrm{phot} \le 0.92 \). 
This range ensures the inclusion of all non-emission-line source candidates within the redshift scope of our study, accounting for the wavelength shift across the field of view and the effective passband width \( \delta \lambda_e \) at the boundaries of the RTF scan (see OTELO-I for details). Additionally, it incorporates the global accuracy limitations ($\pm 0.01$) of the photometric redshift estimations. Applying these criteria, we identified 169 candidates in the OTELO catalogue in this photometric redshift range. These objects also meet the requirement of having photometric data in at least ten observed bands.

With these two samples of ELGs and non-ELGs at $z \sim 0.9$, we then aim to consider only those with FIR emission. With this purpose we cross-match the \otelo\ catalogue with FIR catalogues. \otelo\ field was observed with the Herschel Space Observatory \citep{pilbratt2010herschel} in the PACS Evolutionary Probe (PEP) survey  \citep{lutz11} and the Multi-tiered Extragalactic Survey \citep[HerMES][]{oliver12}, that make use of PACS \citep{poglitsch2010herschel} and SPIRE \citep{griffin2010herschel} instruments. We have re-reduced the images using the latest version of the Herschel Interactive Processing Environment (HIPE)\footnote{\url{https://www.cosmos.esa.int/web/herschel/hipe-download}} and extracted the IR sources using Sextractor \citep{Bertin96}.\\
The PACS images were reprocessed starting from the Level‑1 timelines using a high‑pass–filter (HPF)–based pipeline, following the methodology extensively tested and calibrated by \citet{Popesso2018}. In particular, we modified the standard reduction by adjusting the HPF width and applying dedicated source masking during the filtering step, in order to minimise flux losses associated with the running median removal. Circular masks centred on the source positions were used, and the interpolation of masked samples was enabled to preserve the point‑source signal. These parameter choices are fully consistent with those explored by \citet{Popesso2018}, who demonstrated their impact on the PACS PSF, noise properties, and photometric accuracy at 70, 100, and 160 $\mu$m. The final maps were produced using the photProject task with non‑standard drizzle parameters optimised for point‑source photometry.
The SPIRE maps were reprocessed from the Level‑1 timelines using a map‑making strategy tailored for point‑source analysis, following the methodology described by \citet{Smith2012} and the SPIRE Instrument Control Centre recommendations. Departures from the default pipeline included modifications of the baseline removal and glitch mitigation parameters in order to optimise the recovery of compact sources and minimise residual large‑scale artefacts. In particular, a de-striper-‑based mapper was adopted, with the de-striping polynomial order and masking strategy adjusted following the prescriptions of Smith  who demonstrated the impact of these parameters on point‑source photometry and noise properties at 250, 350, and 500 $\mu$m. These choices have been extensively tested and calibrated in previous Herschel SPIRE analyses and are appropriate for reliable flux extraction of point‑like sources.\\
We have used \textsc{Nway} \citep{salvato18} to cross-match the OTELO catalogue with FIR catalogues. \textsc{Nway} considers spatial proximity and optical photometry (we have used R-band photometry) to calculate the likelihood that sources from different catalogs correspond to the same astronomical object. When a source in a non-optical catalogue has a similar probability to be assigned to two different optical objects, the tool flags it and we then set the corresponding photometric data point as an upper limit  \citep{salvato18}. 

The final sample of emitters and non-emitters observed by \otelo\ with FIR emission at $z \sim 0.9$ is composed of 78 galaxies. All sources are detected either by PACS or SPIRE in at least two bands, ensuring reliable FIR measurements.

\subsection{Line emitters and non-emitters}
\label{ELSs}

The previously mentioned works of \citet{Bongiovanni2020} and \citet{NavarroMartinez2021} analysed the ELGs corresponding to the \oiiis\ and \hb\ lines, respectively. Matching the samples of these works with the preliminary FIR sample obtained in the previous section, we were left with a total of $19$ sources classified as FIR emitters and ELG simultaneously: nine present \hb\ emission, eight sources are \oiiis\ emitters, and two sources present both lines. The preliminary FIR sample with no emission lines sample is hence composed of $59$ objects.

\begin{figure}[ht]
\centering
 \includegraphics[width=0.95\linewidth]{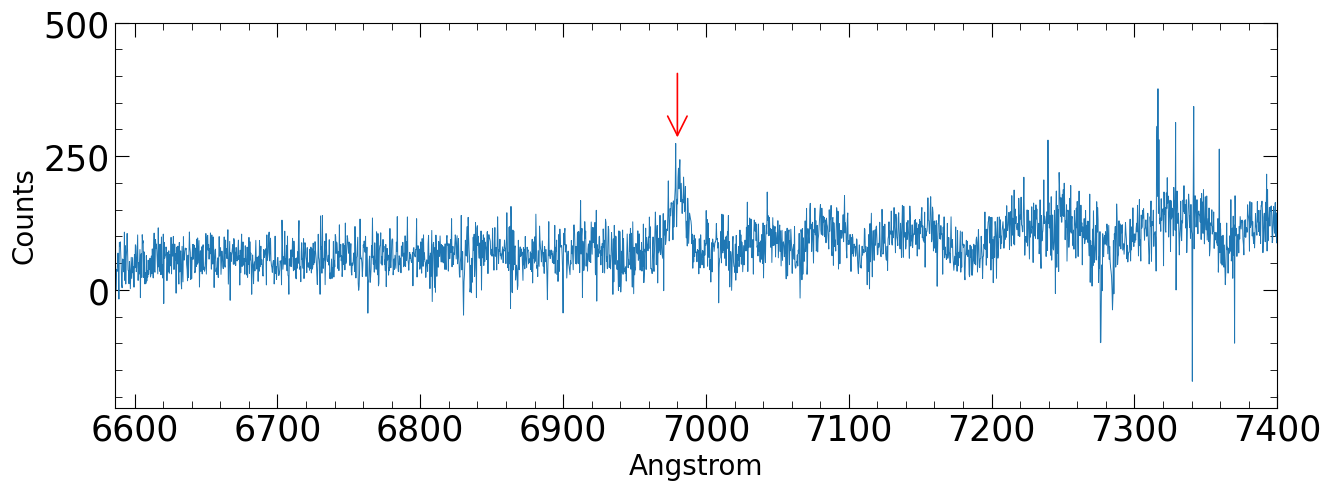}
 \caption{Example DEEP2 spectrum of source 9550, whose emission lines are not detected in the \otelo\ pseudo-spectrum. The \oiis\ line (red arrow) is detected in the DEEP2 spectrum with a very low S/N, indicating intrinsically weak nebular emission. At this redshift, \hb\ and both \oiiis\ emission lines fall within the \otelo\ wavelength range (8950--9300~\AA), but their expected fluxes remain below the \otelo\ detection threshold and may lie close to the edges of the wavelength coverage. Therefore, this source is not classified as an emission-line object in the \otelo\ survey.}
\label{fig:DEEP2spec}
\end{figure}

Among the non-ELGs (noELGs from now on) from the OTELO survey, six have spectroscopic redshift from DEEP2 \citep{newman2013deep2}. From this spectroscopic data we can infer that they are \oii\ line emitters (see example in Fig.~\ref{fig:DEEP2spec}). However, the \oii\ feature in DEEP2 is detected with extremely low S/N, indicating intrinsically weak nebular emission. Under such conditions, the accompanying lines (in particular \hb\ and \oiii) are expected to be even more difficult to detect, either due to their typical relative weakness or because they fall near the edge of \otelo 's wavelength coverage at these redshifts. Hence, these sources are not detected as emitters in \otelo and are treated as non-emitters for the purposes of this work.

\subsection{Identifying AGNs}
Since our interest is exclusively in SFGs, we discriminated between them and AGNs. The first approach to this aim was the use of the X-ray criteria by identifying sources within our sample that show detectable emissions in the X-ray range of the electromagnetic spectrum. The AGNs, including quasars and Seyfert galaxies, are known to emit vast amounts of X-rays due to the high-energy processes occurring in their vicinity, such as accretion onto supermassive black holes. The extended OTELO catalogue provides X-ray data from \textit{Chandra}, reaching a depth of \(\sim 5 \times 10^{-16}\,\mathrm{erg\,s^{-1}\,cm^{-2}}\) in the 0.5--2 keV band \citep{Evans2024}, allowing us to identify X-ray detected AGN candidates at the redshift of our sample. By cross-matching this information with our sample, we identified five X-ray detected AGN candidates.

As an alternative approach, we applied the infrared colour criteria proposed by \cite{Donley2012}, which utilises \textit{Spitzer}/IRAC bands at 3.6, 4.5, 5.8, and 8.0 $\mu$m to define an empirical region predominantly populated by AGNs. Among our sample, two sources satisfy the Donley criteria; however, these correspond to two of the five X-ray AGNs already identified.

The AGN fraction is thus of $\sim$6\%. Four of the sources classified as such were non-emitters and the remaining one is a [\ion{O}{iii}] emitter. These five sources were removed from the sample. Although these galaxies may also host star formation activity, removing this small fraction does not significantly affect our results, while ensuring a SFG sample free of AGN contamination. This leaves a sample of 73 SFGs, 55 of which are noELGs and 18 are ELGs. Notably, after performing the SED fitting with {\sc cigale} \cite[Code Investigating GALaxy Emission;][]{Burgarella2005, Noll2009, Boquien2019}\footnote{{\tt https://cigale.lam.fr}} including an AGN component (see Sec.\ref{sec:SEDfitting}), no galaxy in our sample shows an AGN contribution. This result reinforces the robustness of our initial AGN exclusion criteria and provides additional confidence in the validity of our SFR measurements.

\begin{figure}[ht]
    \centering
    \includegraphics[width=1\linewidth]{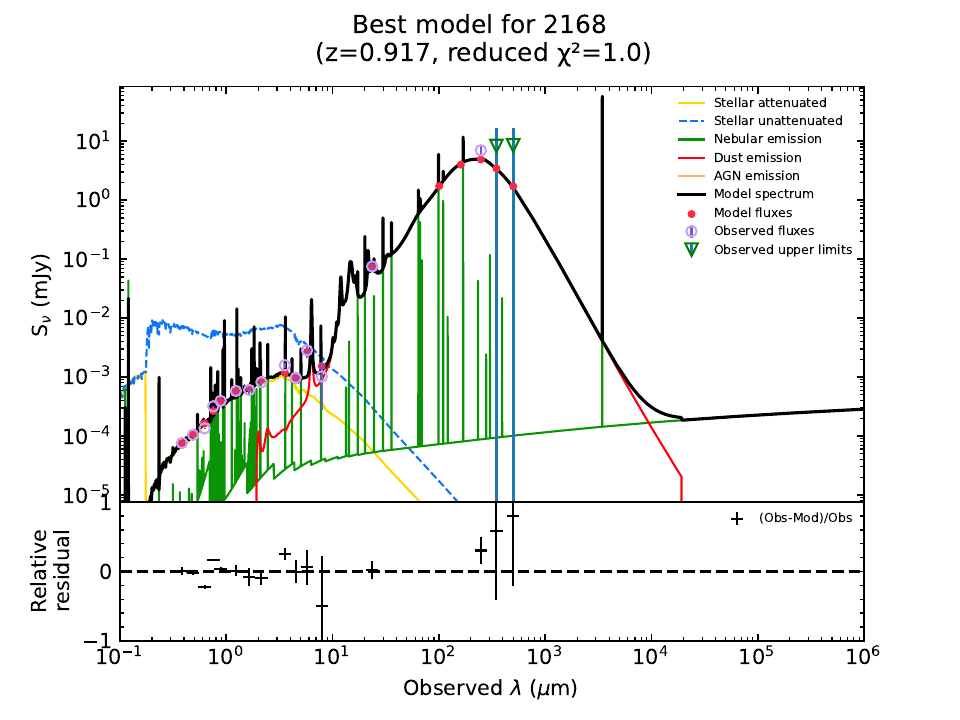}
    \caption{Example of an SED fitting of the galaxy ID2168 produced by {\sc cigale}, which follows the typical tendency of the overall fits. The modules are shown in the legend. Top: Different components. Botom: Relative residuals between the data and the best-fitted model
    }
    \label{SED_example_PDF}
\end{figure}

\section{Spectral energy distribution analysis}
\label{sec:SEDfitting}

The physical properties of the sample (see Sect. \ref{sec:Data}) were calculated by the SED fitting performed with the {\sc cigale} code. This is an SED-fitting code that balances the energy absorbed by the interstellar medium (ISM) in the UV-to-NIR domain with its re-emission in the medium and FIR by the use of different attenuation laws and related parameters. In addition, {\sc cigale} uses a grid of nebular templates to obtain the emission-line spectrum of the system \citep{Inoue2011}. The code requires as input a set of  full wavelength range un-attenuated SEDs, and, after the fitting, it provides a set of physical properties such as SFR, attenuation, dust luminosity, stellar mass, and many other physical quantities. One of the significant advantages of {\sc cigale} over other SED-fitting codes is its ability to include a comprehensive treatment of dust attenuation and re-emission, as well as its flexibility in handling various star formation histories and complex physical processes. 

 The overall quality of the fit is arbitrated by a reduced $\chi ^2$ ($\chi_r^2$). The fitting used is the selected by {\sc cigale} as best model. However, when the $\chi_r ^2$ of the best model is still moderately high ($>2.5$), the fit and the source will be discarded.
\subsection{{\sc cigale} models } 
\label{subsec:cigale-mod}
For the purpose of this work, we fit data up to 18 photometric bands available in the OTELO catalogue. For some sources, measurements in certain photometric bands, particularly in the FIR range, are treated as upper limits, as they may be affected by contamination from nearby sources.

To account for this, we adopt the following criterion: sources for which $25$\% or more of the available photometric data points are upper limits are flagged when deriving their physical parameters with {\sc cigale}. For these sources ($18$\% of the final sample), the resulting physical parameters should be interpreted with caution, as they may be systematically overestimated. We refer to these galaxies as sources affected by FIR upper limits.
 Table \ref{table:bands} summarises the number of bands available for the sample and the specific bands used. Regarding the infrared coverage, nearly 90\% of the sample has detections in at least $7$ IR bands. We assume SSPs with solar metallicity from \cite{Bruzual2003} with a \cite{Chabrier2003} IMF. The SSPs are combined to produce a bursty SFH parametrised as a double exponential (see Sect. \ref{sec:sfh}). We also assume the modified dust attenuation model of \cite{Charlot2000}, the dust and nebular emission model of \cite{Draine2007} and \cite{Inoue2011}, respectively and the AGN component of \cite{Fritz2006}.

\begin{table}
\caption{Bands used for the SED fitting.}            
\label{table:bands}      
\centering                                    
\begin{tabular}{c c}         
\hline\hline                      
Bands available & Sources \\  
\hline                                   
    10 & 1 \\
    11 & 3 \\
    14 & 11 \\
    15 & 37 \\
    16 & 5 \\
    17 & 2 \\
    18 & 14 \\
\hline      
TOTAL & 73 \\
\hline   
\end{tabular}
\tablefoot{List of bands (a maximum of 18): $u$, $g$, $r$, $i$, and $z$ from CFHTLS;  $J$, $H$, and $K_s$ from WIRDS; the four IRAC bands; 24\,$\mu$m from MIPS; 100 and 160\,$\mu$m from PACS and finally 250, 350, and 500\,$\mu$m from SPIRE.}
\end{table}

\subsubsection{Stellar component and IMF}

The choice of IMF and metallicity primarily affects the normalization of the derived SFRs. We adopt a \cite{Chabrier2003} IMF and assume solar metallicity; for discussion of the impact of metallicity on SFR estimates; see \citealp{BBP14} and \citealp{Ceretal16}.

\subsubsection{Star formation history}
\label{sec:sfh}

As a first-order approximation, the SFH of a galaxy can be modelled as a delayed exponentially declining star formation history, which accounts for the bulk of the stellar population \cite[see, among others,][and references therein]{Boquien2019}. However, as we are interested in recent star formation, which accounts for the \hb\ emission present in some of the galaxies of our sample (and the possible obscured star formation in the others), we assumed an additional component of the SFH as well as an exponential, and we let {\sc cigale} decide whether this second component is appropriate in each individual case. In this context, the first exponential describes the enduring star formation responsible for the majority of the galaxy's stellar mass, while the second one characterises a more recent burst of star formation. Since we are interested in the recent star formation (averaged over the last 10 Myrs; see Sec.~\ref{subsec:PhysPropCig}) and given that the star formation is described as an exponential decay, we have use a conservative upper age limit up to 50 Myrs for the most recent burst in such a way that the tail of a burst of 50 Myrs or older will not affect the inference of the recent star formation. The different parameters of the SFH to be fitted are shown in Table~\ref{cigaletable}.

\subsubsection{Dust attenuation}

The \citet{Charlot2000} dust attenuation model is a widely used prescription for the attenuation of stellar light. This dust attenuation combines the birth cloud (BC) attenuation and the interstellar medium (ISM) attenuation, both represented by a power law. 
It makes a distinction between young (age $< 10^7$) and old (age $> 10^7$ years) stellar emission; the young stellar emission is attenuated by the BC whereas the old stellar emission is only affected by the ISM, due to the distinction between the stellar emission components which is a suitable law for IR galaxies \citep{Buat2018}. These two attenuation models are described by a power law and normalised to the amount of attenuation in the V band, $A_V^{\mathrm{ISM}}$ and $A_V^{\mathrm{BC}}$:

\begin{align*}
\centering
\begin{cases}
A_\lambda^{\mathrm{BC}}=A_V^{\mathrm{BC}}(\lambda/0.55)^{n^{\mathrm{BC}}}\\
A_\lambda^{\mathrm{ISM}}=A_V^{\mathrm{ISM}}(\lambda/0.55)^{n^{\mathrm{ISM}}}.
\end{cases}    
\end{align*}

The ratio of the attenuation in the V band experienced by old and young stars, is defined as 
\begin{align*}
\centering
  \mu = \frac{A_V^{\mathrm{ISM}}}{A_{V}^{\mathrm{ISM}}+A_V^{\mathrm{BC}}}.
\end{align*}
We used the modification by \citet{Buat2018}, where they take $\mu$ as a free parameter, since we are interested in allowing some flexibility from the original recipe as we study obscured  star formation. 

\subsubsection{Dust emission}

As dust absorbs stellar photons from the UV to the NIR, it re-emits this energy at longer wavelengths, primarily in the MIR and FIR regions \citep{Calzetti2000, draine2004interstellar}. Dust emission can generally be categorised into three main components. In the MIR region, around 8 $\mu$m, polycyclic aromatic hydrocarbon (PAH) bands dominate the emission \citep{smith2007mid}. At longer wavelengths, very small, warm grains progressively take over the emission, initially undergoing stochastic heating in weak to moderate radiation fields, and eventually displaying equilibrium emission at higher intensities. Beyond approximately 100 $\mu$m, larger, relatively cold grains increasingly contribute to the emission  \citep{Draine2007}. The SED produced by dust is influenced by the different heating mechanisms, dust species, metallicity, and the intensity and shape of the incident radiation field \citep{draine2004interstellar, bianchi2013vindicating}.

One significant advantage of the  model from \citet{Draine2007} lies in its flexibility. It can adopt a wide range of physical conditions, including diverse radiation fields and variable PAH emissions. However, this flexibility results in a considerably larger parameter space to navigate compared to other templates \citep{Boquien2019}.
This model takes into account the emission from small dust grains and the characteristics of intense MIR emission (extreme heating environments tend to stop the process of the dust formation) typically applied for SFGs, which is the case of our sample. The mass fraction of PAHs as well as the minimum radiation field are taken as free parameters. The dust mass fraction linked to the star-forming regions (respectively diffuse emission) is fixed to $\gamma=0.02$ (respectively $1-\gamma=0.98$) for our case.
This model allowed us to determine the dust luminosity of the sample.

\subsubsection{Active galactic nucleus contribution}

Although AGN are identified in our sample using several robust diagnostics, including X-ray criteria, this approach is inherently limited by data availability, as not all galaxies have X-ray coverage. To account for potential residual AGN contamination, we include an AGN component in the SED fitting procedure using the models of \citet{Fritz2006}. In this framework, the AGN contribution is primarily manifested as an additional dust emission component in the MIR, parametrised by the AGN fraction ($f_\mathrm{AGN}$), which represents the fractional contribution of the AGN to the total IR luminosity.
Following previous {\sc cigale}-based studies (e.g. \citealt{ciesla2015constraining}), we do not adopt a strict threshold in $f_\mathrm{AGN}$ to classify galaxies as AGN- or star-formation-dominated. Instead, $f_\mathrm{AGN}$ is used as a diagnostic parameter to assess the potential impact of AGN emission on the derived FIR-based SFRs. The explored range is limited to $f_\mathrm{AGN}<0.25$, as for lower values the AGN contribution to the FIR emission is expected to be small and difficult to constrain individually, while higher fractions would correspond to systems no longer dominated by star formation. Simulations show that for $f_\mathrm{AGN}<0.25$ the effect on FIR-derived SFRs is typically below $\sim$0.1–0.15 dex, particularly for Type-2 AGN \citep{ciesla2015constraining}.
An analysis of the $f_\mathrm{AGN}$ distribution for our sample shows that AGN contributions are generally low. Even in the most extreme cases ($f_\mathrm{AGN}\simeq0.25$), the maximum possible reduction in SFR, assuming the AGN contribution were entirely removed from the total IR luminosity, would be $\lesssim$0.12 dex, well below the typical uncertainties associated with FIR-based SFR estimates. Therefore, the presence of moderate AGN components does not significantly affect our results or conclusions.

\subsubsection{Nebular emission}

The Lyman continuum emitted from the more massive galaxies ionises the surrounding gas which re–emits energy that extends far into the radio regime. This emission is characterised by the presence of Hydrogen and Helium recombination lines, different nebular emission lines related to gas conditions of the emitting gas (temperature and metallicity), and a nebular continuum with extend up to radio waves.  We assume that the nebular lines and continuum are well represented by the used nebular templates from \cite{Inoue2011}. 
They predict the relative intensities of 124 lines from \ion{H}{II} regions from FUV to MIR. 
These templates are parametrised according to ionization parameter $U$ and the metallicity $Z$, which is assumed to be the same as the stellar metallicity ($\mathrm{Z}_\odot$ in this work). 
The electron density is assumed to be constant and is set to $100\;\mathrm{cm}^{-3}$.\\

\subsection{Fitting the SEDs}

The combination of the models used produces $\mathrm{\sim 19}$ million of templates. {\sc cigale} takes into account those templates and computes the SED fits, selecting the best probabilistic result by a Bayesian approach (see \citealt{Noll2009} for details). 
The main parameters of the models described above are summarised in Table \ref{cigaletable}. 

The resulting SEDs (see Fig. \ref{SED_example_PDF}) were visually checked to assess the quality of fit. We found the fits are good for 62 out of 73 sources having a $\mathrm{\chi_r^{2} \sim 1}$ on average and $90\%$ of that sample with a $\mathrm{\chi}_r^{2}$ lower than three. Eleven sources provided an unreliable fitting and high $\mathrm{\chi_r^{2}}$ value and were flagged as Quality $-1$ and removed from the sample.

The final sample of 61 sources (6 \hb\ ELGs, 8 \oiii\ ELGs, and 47 noELGs) is analysed in Sect. \ref{sec:Analysis}. The reason for excluding this additional source is discussed therein.

\begin{figure}[ht]
    \centering
    \includegraphics[width=0.8\linewidth]{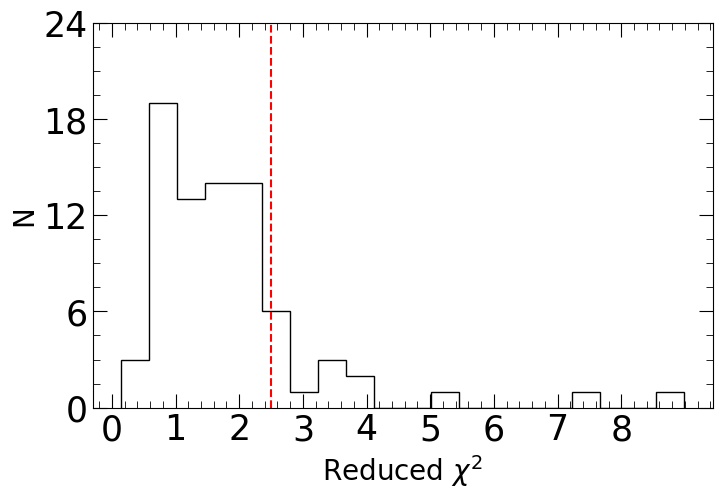}
    \caption{Distribution of the $\mathrm{\chi_r^{2}}$ values obtained for the sample. The dotted vertical red line marks the threshold used to discard poor fits ($\mathrm{\chi_r^{2}}>2.5$).}
    \label{chi2_distribution}
\end{figure}

\begin{table*}[ht]
\caption{Main models and parameters used in {\sc cigale}.} 
\label{cigaletable}  
\centering
\resizebox{\textwidth}{!}{ 
\begin{tabular}[h!]{l l l}
\hline\hline  
\textbf{{\sc cigale} module} & \textbf{Main parameters} & \textbf{Description} \\ 
\hline 
SFH two decreasing   & $\mathrm{f_{burst}=  0.001, 0.05, 0.1, 0.4}$  &  Mass fraction of the late burst population \\ 
exponentials         & $\mathrm{age = 2000, 4000, 5000, 6000, 7000}$ [Myr]    &  Age of the main stellar population in the galaxy  \\
                     & $\mathrm{age_{burst} =  5, 10, 30, 50}$ [Myr]  &    Age of the late burst  \\
                     & $\tau_\mathrm{main} = 1000, 3000, 4000, 6000$ & Width of the main stellar population \\
                     & $\tau_\mathrm{burst} = 10, 50, 100$ & Width of the starburst population\\
                    
 \hline
Single stellar population models & $\mathrm{Z}_{\odot}$ & Metallicity \\                                
\citep{Bruzual2003} & IMF = \citet{Chabrier2003} & Initial mass function \\
\hline
Nebular emission  & $\log U = -2$ & Ionisation parameter   \\
\citep{Inoue2011} &  & \\
\hline
Dust attenuation       &  $\mathrm{Av = 0.5, 0.8, 1.1, 1.4, 1.7, 2.0, 2.3, 2.6 }$ &  V-band attenuation in the birth clouds \\
\citep{Charlot2000}    &  $\mathrm{slope_{BC} = -0.7}$ & Power law slope of the attenuation in the birth clouds \\
                       &  $\mathrm{slope_{ISM} = -1.2, -0.7, -0.4}$ & Power law slope of the attenuation in the  \\
                       & & interstellar medium  \\
 \hline
Dust emision models     &  $q_\mathrm{pah} = 1.12, 2.50, 3.19$   & Mass fraction of PAHs \\ 
\citep{Draine2007}     &  $u_\mathrm{min} = 5, 10, 25$       & Minimum radiation field \\
 \hline

AGN component        & fracAGN = 0.0, 0.1, 0.25  & AGN fraction \\
  \citep{Fritz2006}   & $r_\mathrm{ratio} = 100$   & Opening angle of the dust torus \\ 
                     & $psy = 0.001, 89.990$ & Angle between AGN axis and line of sight \\
 \hline
      \end{tabular}
    }
\tablefoot{The first column lists the {\sc cigale} models, whereas the second column shows the main parameters and the range of values selected for the analysis. The third column provides a brief description of each parameter.}
      
\end{table*}

\section{Analysis}  
\label{sec:Analysis}
The sample of star-forming galaxies with FIR emission, after the exclusion of sources with unreliable SED fits and AGN galaxies, consisted of 62 sources (see Sect. \ref{sec:SEDfitting}), with redshifts spanning from 0.81 to 0.92. 
Since our analysis focuses on the IR regime, we imposed a luminosity threshold of $\log(\mathrm{L}_{IR}/\mathrm{L_\odot}) > 10$. Based on the IR luminosities derived from the SED fitting, one source falls below this threshold and is therefore not included in the final sample analysis. The excluded galaxy is a non–line-emitting source.
In this section, we describe the physical parameters for the 61 sources.

\subsection{Physical properties of the sample from {\sc cigale}}
\label{subsec:PhysPropCig}

To ascertain whether there are significant differences between ELGs and noELGs from our FIR sample (FIRELG and FIRnoELGs from now on; see Table \ref{table:subsamples}), we plotted the histograms of these two subsamples separately. To account for the difference in sample sizes between the two subsamples, the histograms are presented with normalization applied in terms of density. This approach scales the y-axis so that the area under the histogram integrates to one, ensuring that the distributions are represented as probabilities. 

\begin{figure*}
\centering

\begin{minipage}{0.65\textwidth}
\centering
\subfigure{\includegraphics[width=0.48\linewidth]{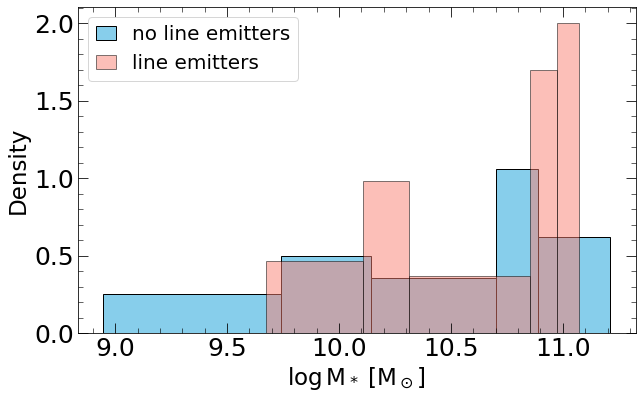}}
\subfigure{\includegraphics[width=0.48\linewidth]{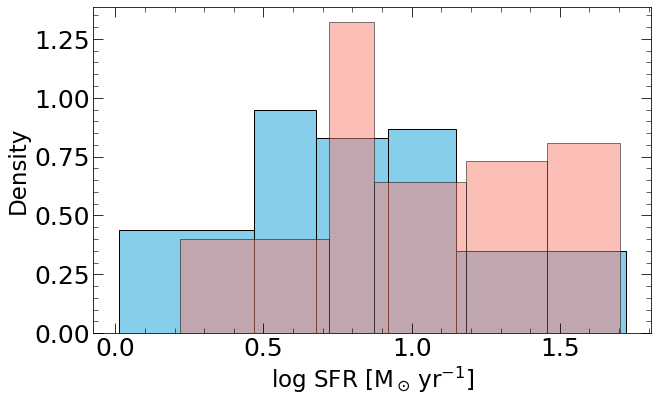}}

\subfigure{\includegraphics[width=0.48\linewidth]{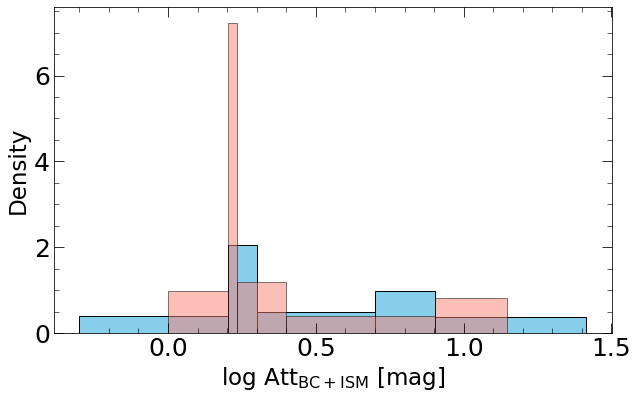}}
\subfigure{\includegraphics[width=0.48\linewidth]{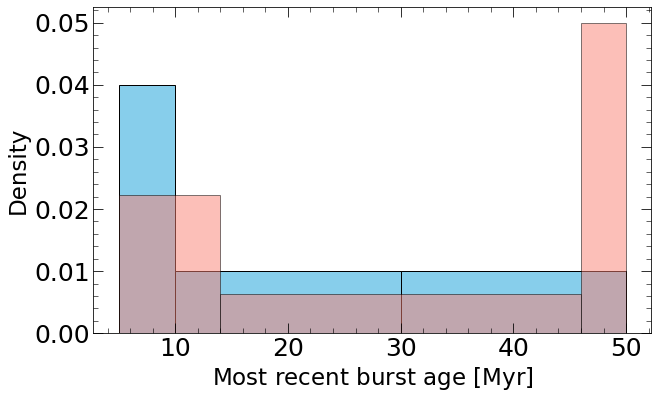}}

\subfigure{\includegraphics[width=0.48\linewidth]{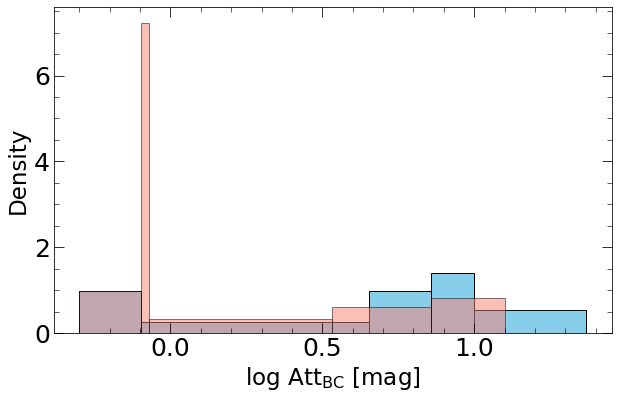}}
\subfigure{\includegraphics[width=0.48\linewidth]{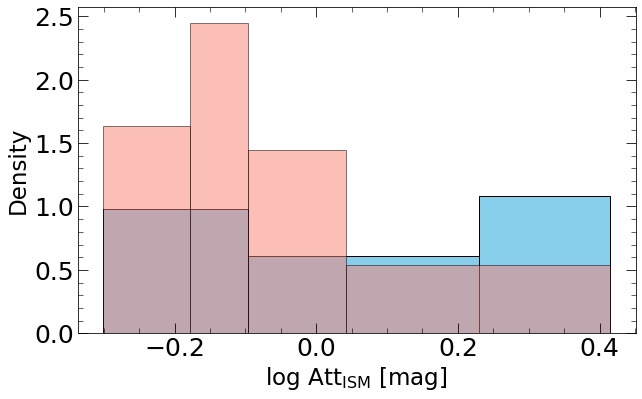}}

\subfigure{\includegraphics[width=0.48\linewidth]{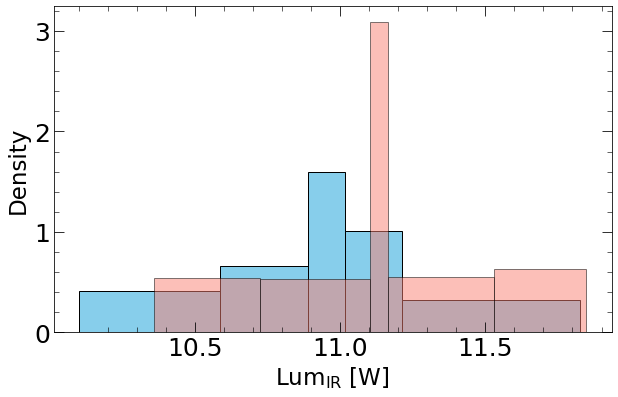}}
\end{minipage}
\hfill
\begin{minipage}{0.30\textwidth}
\caption{Distribution of significant parameters for the sample. The pink histogram represents line-emitting sources, and the blue histogram portrays non-emitters. From top left to bottom right: stellar mass, average SFR over $10\;\mathrm{Myr}$, attenuation, age of the most recent starburst, birth-cloud attenuation, interstellar medium attenuation, and IR luminosity. Each bar represents an equal probability (20\%, except 33\% for burst age).}
\label{fig:histograms}
\end{minipage}

\end{figure*}

Stellar masses are derived by the \cite{Bruzual2003} module in {\sc cigale}. This module discriminates between the stellar masses of young and old populations. When referring to stellar mass, however, we consider the addition of these two. The stellar mass spans the range $8.7 \le \log M_\star/\mathrm{M}_\odot \le 11.2$, with a typical uncertainty of $\pm0.24$ dex. The distribution of $\log M_\star$ is shown in Fig. \ref{fig:histograms}.

The SFR provided by the {\sc cigale} module, assuming an SFH of two exponentials, calculates an estimation of the average SFR over the last $10\; \mathrm{Myr}$ as well as the average SFR over $100\; \mathrm{Myr}$ and the instantaneous SFR at the age of the galaxy.  For this work, we adopt the SFR averaged over the last $10\; \mathrm{Myr}$ as it provides recent star formation activity in the galaxy and allows for comparison with other SFR tracers. This parameter ranges between $1$ and $53\;\mathrm{M}_\odot\,\mathrm{yr}^{-1}$, as shown in Fig. \ref{fig:histograms}. The distribution is fairly uniform, with a slight tendency towards intermediate values. The typical uncertainty in the SFR estimates is approximately $3.8~\mathrm{M}_\odot\,\mathrm{yr}^{-1}$, with individual errors ranging from $0.4$ to $10.9~\mathrm{M}_\odot\,\mathrm{yr}^{-1}$, as derived from the fits. Differences between the two subsamples are explored further in Sect. \ref{sec:Discussion}.  
A 40\% of the galaxies exhibit SFRs exceeding 
$10\;\mathrm{M}_\odot\,\mathrm{yr}^{-1}$, reaching a 64\% when considering only the FIRELG sample. The FIRnoELG population with SFR over $10\;\mathrm{M}_\odot\,\mathrm{yr}^{-1}$ is 32\%, revealing an outstanding global star-forming activity. We note that In Fig.~\ref{fig:histograms} there is a peak at 50 Myrs for the most recent burst age. However, such peak has not a physical meaning and it is an artefact of the choice of the maximum burst age of the recent component (see also Sec.~\ref{sec:sfh}).

The {\sc cigale} module \cite{Draine2007} estimates the dust luminosity, accounting for the IR luminosity, which, in the case of this sample, ranges from $10.1 \le \log L_\mathrm{IR}/\mathrm{L}_\odot \le 11.8$ with an average uncertainty of $0.49$ dex. Figure \ref{fig:histograms} shows the distribution of this parameter.
We find that roughly $50\%$ of the sample qualifies as luminous infrared galaxies, $70\%$ of which are galaxies with no emission lines. 
The attenuation distribution for our sample, shown both in general and separated by the BC and ISM components, demonstrates that most galaxies have moderate to low attenuation values. The BC attenuation of ELGs peaks at lower values compared to ISM attenuation, consistent with the expectation that for ELGs the BC typically contributes less to the total attenuation than the diffuse ISM allowing emission lines to be detectable. 
This distribution provides insights into the role of dust in obscuring star-forming regions across the sample, as FIRELG exhibit lower attenuation overall, while FIRnoELG experience significantly higher levels of dust obscuration. This pattern reflects differences in star-forming environments, where emission-line galaxies are characterised by more active, optically accessible star-forming regions, whereas non-line emitters harbor star formation primarily hidden by dust. 
For the total (BC plus ISM $A_\lambda$) dust attenuation, both distributions are quite flat, with small peaks below $1$ magnitude. 

\subsection{Physical properties of the sample from previous works}
\label{subsec:PhysPropComp}

\begin{figure}[ht]
\centering
\subfigure{\includegraphics[width=0.48\columnwidth]{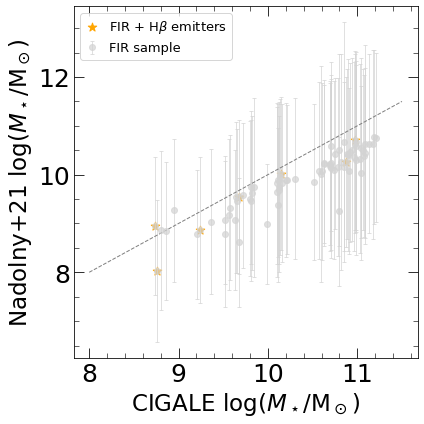}}    \subfigure{\includegraphics[width=0.485\columnwidth]{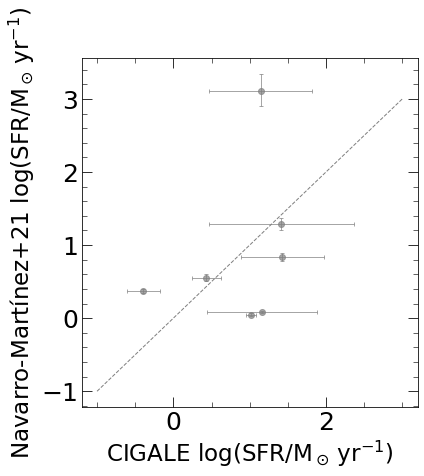}}
 \caption{Comparison of the physical properties derived in this work with estimates from previous studies for sources in common. 
\emph{Left:} Stellar mass comparison between values obtained with {\sc cigale} and those derived by \citet{nadolny2020otelo} using the prescription of \citet{lopez2019alhambra}. 
\emph{Right:} Comparison between SFRs derived with {\sc cigale} (10 Myr timescale) and those obtained from the H$\beta$ line flux in \citet{NavarroMartinez2021} for the six sources that are both FIR and H$\beta$ emitters. In both cases the dashed line indicates the one-to-one relation and error bars represent the uncertainties reported in the respective works.} 
\label{fig:mass-sfr-comp}
\end{figure}

Before moving on to the discussion it is worth mentioning that for some of the sources of our whole sample, we have complementary estimations of physical properties such as stellar masses and SFR. \citet{nadolny2020otelo} estimated stellar masses for the full \otelo\ catalogue using the recipe described in \citet{lopez2019alhambra}. The left panel of Fig. \ref{fig:mass-sfr-comp} shows the comparison between the mentioned estimation and that from {\sc cigale}, portraying an agreement that further validates the reliability of the parameters derived. 

For the SFR, we take advantage of the estimations done in \citet{NavarroMartinez2021} for the\hb\ emitter sample from \otelo\, derived from the \hb\ flux. These SFRs were obtained following the \citet{Kennicutt2012} calibration adapted to \hb, assuming a \cite{Kroupa2001} IMF and correcting for dust extinction using the case B recombination ratio $I(\mathrm{H\alpha})/I(\mathrm{H\beta}) = 2.86$ \citep{Storey1995}. Although the SFR of the \hb\ sample is computed assuming a Kroupa IMF and the SFR from the FIR sample is derived with {\sc cigale} assuming a Chabrier IMF, the difference between these two IMFs is very small (less than 0.05 dex; e.g. \citealt{madau2014cosmic}) and well below the typical uncertainties of both methods. Therefore, no IMF rescaling is applied.
In this case, the comparison can be done only for the $6$ sources that are simultaneously \hb\ and FIR emitters. When comparing the SFR estimations by {\sc cigale} with those obtained in \cite{NavarroMartinez2021} using the \hb\ line flux, it is appropriate to use the $10\;\mathrm{Myr}$ SFR since \hb\ emission accounts for recent star formation \citep{Ceretal16}. As shown in Fig. \ref{fig:mass-sfr-comp}, the estimations for the \hb-FIR differ from those presented in \cite{NavarroMartinez2021} for the very nature of these two calibrators. This topic is deeper elaborated in Sect. \ref{subsec:ir-tracer}.

\section{Discussion: The hidden SFR}
\label{sec:Discussion}

In this section we analyse the obscured component of the star formation activity in our FIR-emitting galaxies and compare it with the properties of the non-FIR \hb\ emitters from \citet{NavarroMartinez2021}. This allows us to assess the fraction of star formation that remains undetected when relying solely on optical tracers.

\subsection{Infrared emission as an SFR estimator}
\label{subsec:ir-tracer}
\begin{figure}[ht]
\centering
 \subfigure{\includegraphics[width=0.4\textwidth]{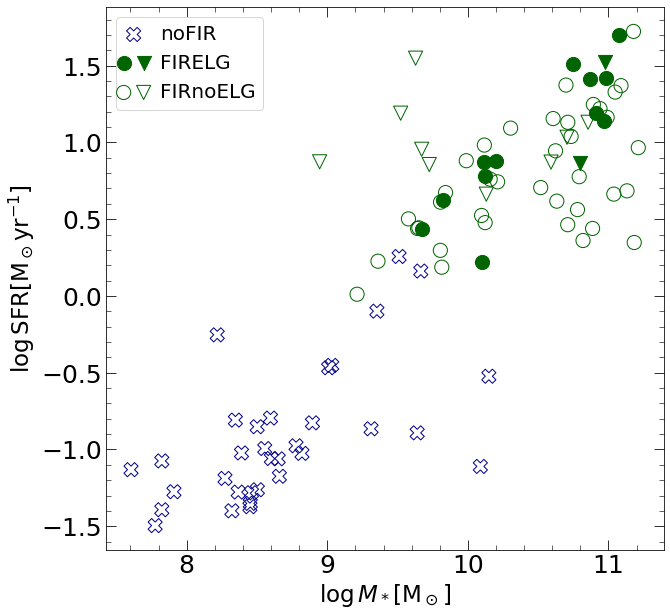}}
 \subfigure{\includegraphics[width=0.4\textwidth]{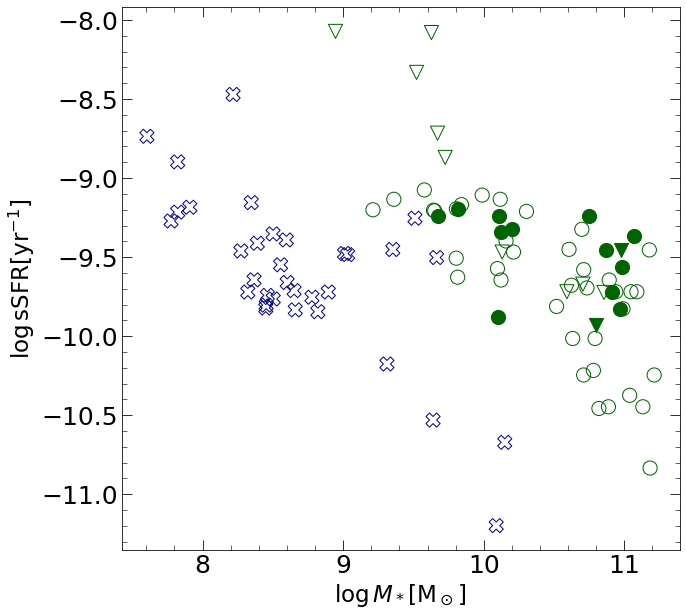}}
 \caption{Mass versus SFR of galaxies at $z \sim 0.9$. This work's whole FIR emitting sample (green circles and triangles) and non-FIR emitters from the \hb\ OTELO sample of \cite{NavarroMartinez2021}, represented by empty crosses. Filled circles or triangles represent FIRELGs, and empty markers are FIRnoELGs. Downward triangles indicate sources affected by FIR upper limits (as elaborated in Sect. \ref{subsec:cigale-mod}), which could result in an overestimation of the plotted parameters.}
 \label{fig:FIRvsnoFIR}
\end{figure}

Here we compare the SFR and stellar mass values of the FIR and noFIR samples described in \ref{table:subsamples}. For the FIR sample, both the stellar masses and SFRs are taken from the {\sc cigale} fits, with the SFR specifically adopted as the average over the last $10\;\mathrm{Myr}$ (SFR$_{10\;\mathrm{Myr}}$), providing a robust tracer of recent star formation activity. In contrast, for the noFIR sample, the stellar masses are taken from \citet{nadolny2020otelo}, derived using the recipe described in \citet{lopez2019alhambra}, while the SFRs are those derived in \citet{NavarroMartinez2021} from the \hb\ flux. This approach ensures a consistent comparison between the two samples, with the FIR SFR$_{10\;\mathrm{Myr}}$ representing ongoing star formation and the \hb-based SFR tracing the same recent timescale. The {\sc cigale} $M_\star$ values for the FIR sample tend to be higher than those reported by \citet{NavarroMartinez2021}. The distribution shown in Fig. \ref{fig:histograms} indicates that our sample includes a substantial fraction of massive galaxies with $\log M_\star > 10.5$, highlighting a systematic shift towards higher stellar masses compared to the \hb\ emitter population, where most masses fall below $10^{9.5}$. The stellar masses and SFR are plotted for both samples in Fig. \ref{fig:FIRvsnoFIR}.  This figure shows the noFIR population located in a lower mass range and hence a lower SFR due to a combination of factors related to ionization, dust, and the detection limitations of the emission lines.

First, young hot stars that ionise H~{\sc ii} regions in galaxies have an upper limit to the amount of ionising radiation they can produce. Specifically, the emission from these regions is capped at around $10^{49}\;\mathrm{erg~ s^{-1}}$, which corresponds to the maximum luminosity of photons a typical H~{\sc ii} region can emit \citep{kennicutt1998star}. This ionization process also has a finite duration, usually lasting on the order of 5 million years \citep{villaverde2010modeling}. 

On the other hand, the majority of the FIR sample ($\sim 80\%$) are disc-like galaxies following the morphological classification detailed in \cite{nadolny2020otelo}, and larger late-type galaxies tend to have more massive stellar populations and, consequently, more dust. 
The ionising radiation from star-forming regions is absorbed by dust and re-radiated in FIR. This attenuation makes it harder to detect \hb\ emission in galaxies with higher stellar and dust masses. As a result, even the larger galaxies, which host more massive stellar populations, may not be detected as ELGs. 
Simultaneously, the more massive a disc galaxy is, the more dust it tends to contain, leading to stronger FIR emission. Given a higher continuum emission, a spectral line of fixed flux exhibits a smaller equivalent width (EW), thereby becoming more difficult to detect. All of this explains why the FIRnoELG sample is predominantly composed of more massive galaxies with significant dust content.

In contrast, star formation in our low-mass galaxies is more easily detected through \hb\ emission. These galaxies typically have less dust and fewer ionising regions, but when they do have ionising regions, the emission lines are less affected by dust attenuation. This is due, as mentioned before, to the BC playing a less significant role in the attenuation than the global ISM. Additionally, the \otelo\ survey in particular is designed to detect emission lines with a low EW, and hence it is biased towards low-mass galaxies, as seen in \cite{cedres2021otelo}. 

Thus, the difference in mass between the FIR and noFIR samples is a direct result of how dust affects the detection of emission lines in these galaxies. While the FIR sample is dominated by more massive dust-rich galaxies that emit strongly in the infrared, the noFIR sample includes smaller, less dusty galaxies where the line emission can be more easily observed.

As noted in the introduction (Sec.\ref{sec:intro}), FIR emission traces the star formation activity of the same young stellar populations responsible for nebular emission lines \citep[e.g.][]{kewley2002halpha}. This provides a physical basis for the comparisons and discussion between FIR- and H$\beta$-selected galaxies.

 \subsection{Star formation rate in obscured galaxies}
\label{subsec:obscuredSFR}

Once the difference between optical and infrared tracers has been established, we now analyse the quantitative impact of the obscured SFR. We address the eventual discrepancies between those galaxies presenting an \hb\ or \oiiis\, and those without an emission line detectable by \otelo ~. 
As mentioned in Sect. \ref{sec:Data}, six sources from our sample are \oii\ emitters as confirmed by the DEEP2  spectra available \citep{newman2013deep2}. These galaxies may still host significant star formation heavily obscured by dust,
therefore lying below the \otelo\ detection threshold. Hence, these galaxies are classified as non-emitters in this work (see Fig. \ref{fig:DEEP2spec}).

\begin{table*}
\caption{Summary of the subsamples used in this work.}
\centering 
\begin{adjustbox}{max width=0.99\textwidth,center}
\begin{tabular}{l c l}
\hline\hline
Shorthand & Sources  & Description\\ 
\hline
FIRnoELG & 47 & Sources in the \otelo\  sample presenting FIR emission but not line emission\\
FIRELG & 14 & Sources in the \otelo\  sample presenting FIR emission as well as line emission\\
FIR & 61 & Whole sample of \otelo\  sources with FIR emission\\
noFIR  & 33  & Sample of \otelo\  sources at $z\sim 0.9$ with \hb\ emission but no FIR emission\\
\hline
\end{tabular}

\end{adjustbox}
\tablefoot{Note that FIR sample is composed by the addition of FIRELGs and FIRnoELG.}
\label{table:subsamples}
\end{table*}

Figure \ref{fig:histograms} shows the distributions of the main parameters obtained from the {\sc cigale} SED fitting, discriminating in each case the FIRELG subsample from the noELG counterpart. For all parameters, the distributions of these two populations are very similar, suggesting that these two populations are not inherently different.

\begin{figure*}[ht]
\centering
\subfigure{\includegraphics[width=0.475\textwidth]{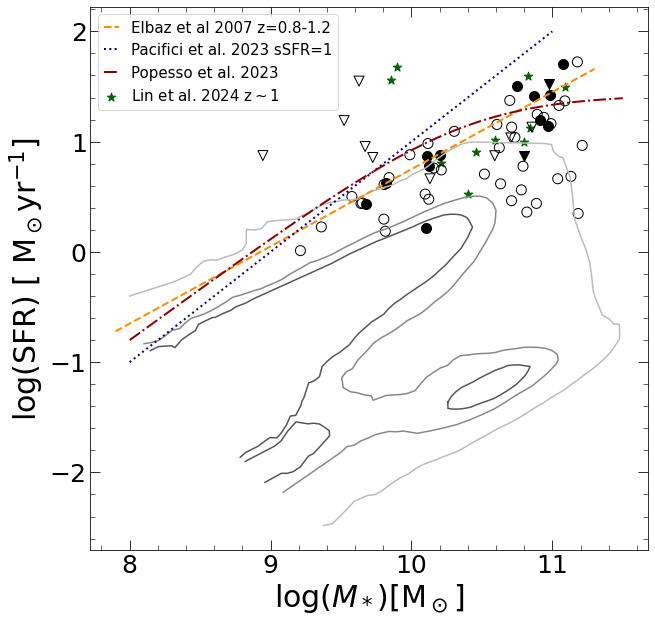}}    \subfigure{\includegraphics[width=0.505\textwidth]{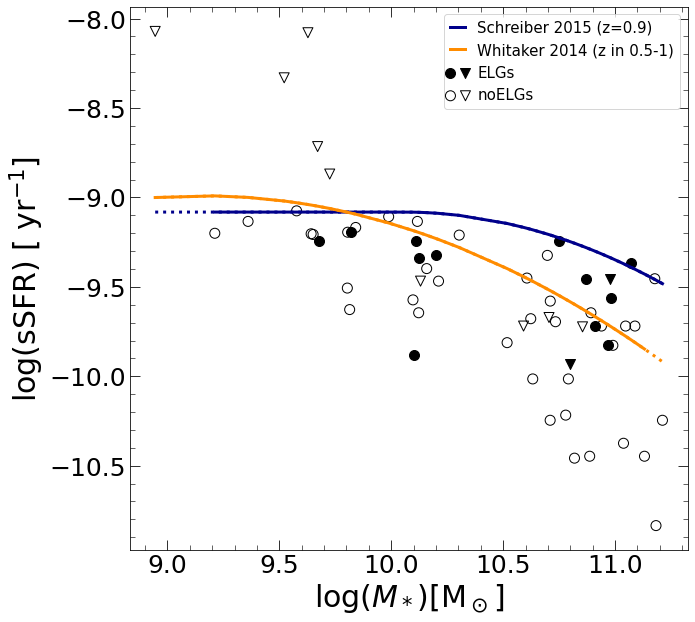}}
 \caption{Star formation rate as a function of $M_\star$ (top) and sSFR (SFR divided by stellar mass, bottom) for the whole sample. Line emitters and non-emitting galaxies are indicated with filled and unfilled markers, respectively. Downward triangles mark those sources affected by FIR upper limits. For the plot on the left, contours correspond to the number density of galaxies from the SDSS database and those obtained by \cite{renzini2015objective} at values of $1.2 \times 10^5$ (inner), $7.0 \times 10^4$ (middle), and $2.0 \times 10^4$ (outer). These contours reveal the position of the star formation MS for local galaxies. Green stars represent the PAH luminous galaxies at $z\sim 1$ sample from \cite{lin2024polycyclic}. The plot on the right shows the different positions of the star formation MS given by \cite{schreiber2015herschel} and \cite{whitaker2014constraining}. For each MS, we differentiate the mass range used by each author from the extrapolation by plotting the latter with a discontinuous line.}
\label{fig:sampleSFR}
\end{figure*}

Figure \ref{fig:sampleSFR}(a) shows the stellar mass as a function of the SFR. Coloured contours correspond to the number density of galaxies from the SDSS database and obtained by \cite{renzini2015objective}, showing the position of the star formation MS for local galaxies.
Our galaxies lie systematically above the contours of local galaxies, indicating that they exhibit higher SFRs at fixed stellar mass. This offset is interpreted as a reflection of both redshift evolution of SFR together with the evolution of the processes involved in star formation activity.   
At $z\sim0.9$, galaxies are expected to exhibit higher SFRs due to the increased availability of cold gas and the enhanced efficiency of star formation processes in the earlier universe. This is consistent with the well-established redshift evolution of the star-forming MS \citep{noeske2007star}.

Additionally, the presence of FIR emission in our sample suggests that many of these galaxies are undergoing starburst episodes. The higher SFR for starburst galaxies places them about the MS. The simultaneous detection of FIR and line emission in some of our sources provides further evidence of active star-forming regions with substantial dust attenuation.

The location of our sample in this diagram is consistent with the MS of \cite{elbaz2007reversal}, who derive the SFR from FIR luminosity using the \cite{kennicutt1998star} recipe of a sample of ultradeep imaging at $24\mu m$ with MIPS at $0.8 \le z \le 1.2$ to determine the contribution of obscured light to to star formation. The dotted line of this figure  represents a constant specific star formation rate (sSFR) equal to 1, as portrayed in \cite{pacifici2023art} with a sample of galaxies at $z\sim 1$. 
This underscores that the majority of our sample follows the expected sSFR for star-forming galaxies, despite exhibiting higher total SFRs than local galaxies. Our sample is still part of the broader population of star-forming MS galaxies at this epoch ($z\sim1$), with a significant number of sources clustering near sSFR$=1\;\mathrm{Gyr}^{-1}$. 
Then, the elevated SFRs observed in our sample, relative to local galaxies, is consistent with redshift evolution

Regarding Fig. \ref{fig:sampleSFR}(b), our sample is consistent with the MSs for sSFRs by \citet{whitaker2014constraining} and lies slightly below \cite{schreiber2015herschel}. This result is consistent with the characteristics of our sample and the physical processes governing star formation in massive dust-rich galaxies. Our sample is selected based on FIR emission, which typically traces dust-obscured star formation. While the FIR provides a robust measure of the total SFR, the galaxies in our sample may represent a population where star formation is heavily enshrouded by dust, potentially leading to systematically lower sSFRs compared to UV/optical-selected samples that dominate the derivation of the MSs. Fundamentally, FIR-bright galaxies tend to have higher stellar masses and lower sSFR compared to UV/optical-selected galaxies. This characteristic aligns with the nature of our sample, where higher mass correlates with lower normalised star formation efficiency. This could explain why our galaxies fall slightly below other populations.

The green star markers in that same plot represent a sample of $10$ PAH luminous galaxies at $z\sim 1$ by \citet{lin2024polycyclic}. Since these galaxies are IR-bright and at a similar redshift to our sample, their position on the SFR–$\mathrm{M}_\star$ plane provides a meaningful comparison. The strong agreement between the two samples further reinforces the conclusion that FIR-detected galaxies predominantly reside along the MS of star-forming galaxies at this epoch.

Higher-mass galaxies in the universe at $z\sim0.9$ are known to exhibit declining sSFRs as they approach the upper envelope of the MS \citep{whitaker2014constraining}. This trend reflects the onset of processes such as gas depletion or feedback from AGNs that regulate star formation. Given that our sample consists of FIR-bright galaxies, which are often associated with higher stellar masses, it is reasonable for their sSFRs to cluster below the MS at this redshift, reflecting the reduced star formation efficiency typical of massive systems.

\subsection{Estimation of total SFR and co-moving density}
\label{subsec:totalSFR}
Building on the quantitative estimate of the obscured star formation derived in the previous section, we now examine this hidden component. We assess how it shapes the global properties of the galaxy population at $\mathrm{z}\sim0.9$.

\begin{figure}[ht]
\centering
 \includegraphics[width=0.9\linewidth]{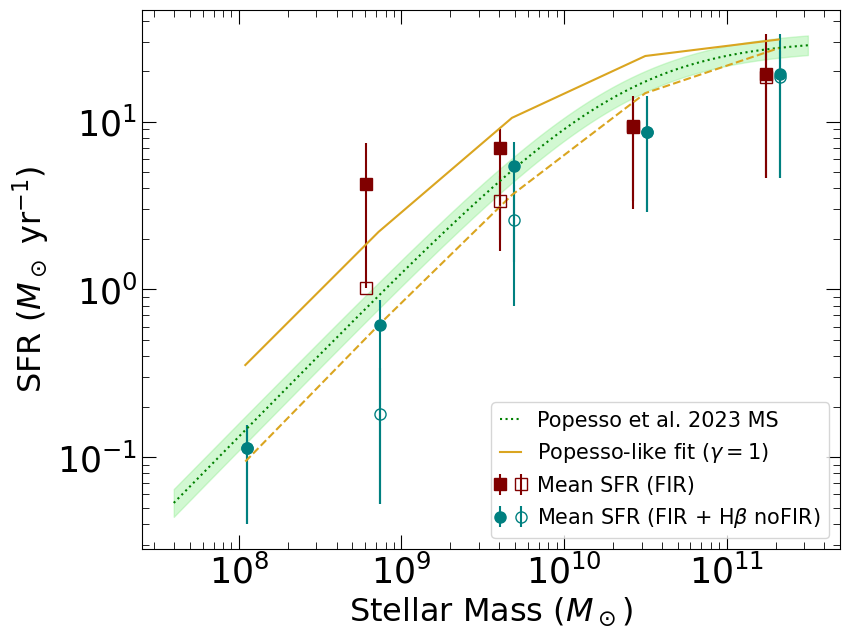}
 \caption{Mean SFR per stellar-mass bin of equal logarithmic width for the FIR-emitter sample (squares) and the combined FIR plus H$\beta$ no-FIR sample (circles) using identical mass bins for both. The number of galaxies per bin for the FIR-only sample is [2, 14, 12, 29, 4], and for the combined sample, it is [11, 17, 20, 30, 16]. Filled symbols include sources affected by FIR upper limits, while open symbols show the results after excluding them. Error bars indicate the 68\% central interval of the SFR distribution within each bin. The dotted green line and contours represent the star-forming MS $\pm1\sigma$ at $z\sim0.9$ from \citet{popesso2023main}. The solid orange line shows a Popesso-like fit to the FIR-only binned data, with $\log(\mathrm{SFR}_{\max}) = 1.51$ and $\gamma = 1$ fixed, following the parametrization in Table~2 of \citet{popesso2023main}. The dotted orange line shows the same fit after excluding sources affected by upper limits.}
 \label{fig:sfr-per-bin}
\end{figure}

Figure \ref{fig:sfr-per-bin} shows the average SFR per stellar-mass bin for the combined samples of FIR galaxies (with and without emission lines) and ELGs without FIR emission. This enables a more precise look at how star formation varies with galaxy mass. Two versions of the relation are shown: one including all FIR emitters and one excluding sources affected by FIR upper limits, as defined in Sect. \ref{subsec:cigale-mod}.

The SFR reveals a clear trend with stellar mass: higher-mass galaxies ($M_\star \ge 10^{10}~\mathrm{M}_\odot$) exhibit elevated SFRs, highlighting their dominant role in driving star formation activity during this epoch. To further interpret our results, we included in Fig.~\ref{fig:sfr-per-bin} the MS relation at $z=0.9$ from \citet{popesso2023main}, which provides an empirical parametrization of the SFR--$M_\star$ relation across cosmic time. This relation follows a smoothly rising trend that flattens at high stellar masses, capturing the so-called turnover mass ($\mathrm{M}_0$). This parameter marks the stellar mass at which the number of SFGs stops rising and begins to decline, defining the bending point in the star formation MS and separating galaxies with nearly constant sSFR (below) from those where star formation is suppressed (above). The turnover mass reported by \citet{popesso2023main} at $z\sim0.9$ is $\log \mathrm{M}_0 = 10.39 \pm 0.05$.

In addition, we performed a fit to our binned data using a Popesso-like functional form (see Eq.~14 in \citealt{popesso2023main}), fixing $\log \mathrm{SFR}_{\max} = 1.51$, corresponding to the expected value at $z=0.9$ from their Table~2, and setting the slope parameter $\gamma = 1$. The fit was also performed both including and excluding sources affected by FIR upper limits, yielding consistent results. The resulting fit, also shown in Fig.~\ref{fig:sfr-per-bin}, provides a turnover mass of $\log \mathrm{M}_0 = 10.55 \pm 0.10~\mathrm{M}_\odot$, in good agreement with the value from \citet{popesso2023main}. The impact of FIR upper-limit–affected sources is confined to the lowest stellar-mass bin ($\sim10^9~\mathrm{M}_\odot$) and does not affect the overall trends or conclusions of the analysis.

Figure \ref{fig:sfr-per-bin} also manifests the need of considering SFRs estimated from both emission line and $L_{IR}$ tracers.
While the higher end of the stellar mass range is not affected when using only the $L_{IR}$ indicator, the low-mass end ($\le 10^9~\mathrm{M}_\odot$) shows the impact of not including the SFR from emission line galaxies without IR emission, where the low-intermediate mass galaxies contribute significantly.
In this way, the apparent excess of SFR with respect to that expected from the MS disappears when including also the contribution from the emission line indicators. At the same time, it shows a very good match with the MS in the low-mass range.
Our sample is therefore formed by galaxies that follow the evolutionary path of the MS at this redshift range.

Table \ref{table:sfr-density} shows the SFR density (SFRD) within the co-moving volume of the global sample, which is estimated to be approximately $1.25 \times 10^4 \pm 1.59 \times 10^3\ \mathrm{Mpc}^3$. The volume is constrained by the redshift range of the full FIR sample ($0.81 \le z \le 0.92$). The SFRD is computed by summing the SFRs of all galaxies in the sample and dividing by the co-moving volume, with uncertainties propagated from individual SFR errors.
 
As anticipated, the SFRD obtained from the whole FIR sample is significantly higher than that derived only from the emission-line subsample. This shows that non-ELGs contribute significantly to the total SFRD, reinforcing the importance of including obscured star formation. 
The factor of three difference between the SFRD of ELG alone and the total sample underscores how, as argued before, relying solely on optical surveys underestimates the true star formation activity, particularly in dusty, FIR-emitting galaxies. The mean SFR per galaxy is higher for ELG. This suggests that while ELG galaxies individually form stars at a higher rate, the larger population of non-ELGs compensates for their lower individual SFR, leading to a higher cumulative contribution to the total SFRD. This trend also reflects the typically lower stellar mass of ELGs, highlighting the key role of optical tracers in identifying and accounting for star formation in these low-mass systems.

\begin{figure}[ht]
\centering
 \includegraphics[width=0.4\textwidth]{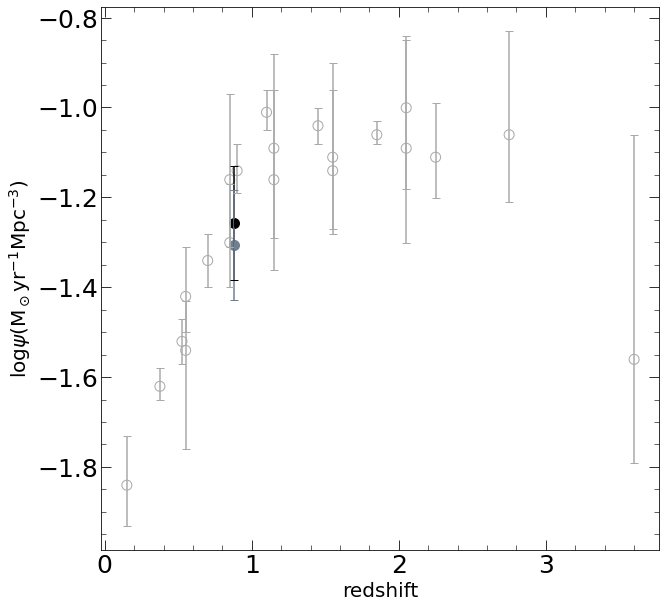}
\caption{Star-formation rate density as a function of redshift. Grey empty markers show the compilation from \citet{madau2014cosmic}, adjusted to a Chabrier IMF. The black filled circle shows the SFRD derived for our FIR-emitter sample including sources affected by FIR upper limits, while the black open circle shows the SFRD obtained after excluding these sources.}
 \label{fig:madau}
\end{figure}

Figure \ref{fig:madau} shows the evolution of the SFRD derived from FIR emission as a function of redshift. We compare our result with those reported by \cite{madau2014cosmic}, converted to a Chabrier IMF. 
Two estimates of the SFRD are shown: one including all FIR emitters and one excluding sources affected by FIR upper limits, as defined in Section~X.
Our estimate of the SFRD is consistent with values reported in the literature within the uncertainties.  Including sources affected by FIR upper limits yields an SFRD closer to the Madau--Dickinson relation, while excluding them provides a more conservative lower bound. The true value is therefore expected to lie between these two estimates.
However, the potential impact of cosmic variance (CV) remains a relevant consideration, given the limited volume probed by the \otelo\ survey. Its ultra-deep, pencil-beam design enhances sensitivity to faint emitters, especially low-mass galaxies (OTELO-I), but at the cost of sampling a narrow field that may not fully capture the large-scale structure of the universe. This increases susceptibility to CV, which can introduce significant statistical uncertainty in derived quantities such as the SFRD.
 Previous studies estimate that surveys with areas comparable to OTELO's may experience CV on the order of 20–30\% \citep{somerville2004cosmic}. Including this contribution would not significantly alter our conclusions but is important for context when comparing with results from wider-field surveys.

\begin{table*}
\caption{Star formation metrics for each subsample.}
\label{table:sfr-density}     
\centering                                      
\begin{tabular}{c c c c c}       
\hline\hline
&FIRELG& FIRnoELG& NoFIR & Total\\     
\hline                                   
Number of galaxies & 14 & 47 & 33 & 94
\\
Gross SFR ($\mathrm{M}_\odot \mathrm{yr}^{-1}$) & $234.71$ & $447.36$  & $7.83$ & $689.9$ \\ 

SFRD ($\mathrm{M}_\odot \mathrm{yr}^{-1} \mathrm{Mpc}^{-3}$)  & $1.88\times 10^{-2}$ & $3.58\times10^{-2}$ & $6.27\times 10^{-4}$& $5.52\times10^{-2}\pm 2.54\times10^{-3}$ \\ 

sSFRD ($\mathrm{yr}^{-1} \mathrm{Mpc}^{-3}$)  & $4.24\times 10^{-13}$ & $3.03\times10^{-12}$ & $1.2\times10^{-12}$&  $4.66\times10^{-12}\pm 3.87 \times 10^{-13}$\\ 

SFR per galaxy ($\mathrm{M}_\odot \mathrm{yr}^{-1})$  & $16.8$ & $9.52$ & $0.24$ & $11.44$\\
\hline                                             
\end{tabular}
\tablefoot{Gross SFR, SFRD, sSFR density (sSFRD), and SFR per galaxy for the FIR line emitting subsample, FIR non line-emitting subsample, the ELG non-FIR emitting subsample, and the compound of all these. The co-moving volume of for the redshift range is $12481.25\;\mathrm{Mpc}^3$.}
\end{table*}

\section{Conclusions}
In this work, we have conducted a detailed analysis of the star formation activity in galaxies at $z \sim 0.9$ using FIR data complemented with deep optical spectroscopy. Beyond computing the cosmic SFRD and the SFR to stellar mass relation, our results provide a more nuanced understanding of the galaxy populations contributing to cosmic star formation at this redshift, with particular focus on the obscured regime.

We found that FIR-based SFRs are broadly consistent with H$\beta$ estimates corrected for extinction, so they provide robust estimates of the total star formation, especially in dust-obscured galaxies. Nevertheless, optical tracers remain essential for identifying star formation in low-mass or less obscured systems not detected in the FIR. 

Overall, this study demonstrates the need for a multi-wavelength approach to fully capture the complexity of star formation across galaxy populations. The FIR and optical tracers play complementary roles: The former excels in massive, dust-obscured galaxies, while the latter is crucial to detecting star formation in low-mass, less obscured systems. These results emphasise the importance of combining IR and spectroscopic data to obtain a complete and accurate view of star formation.

We analysed the SFRD contribution as a function of stellar mass and emission-line properties. Galaxies without emission lines, despite lower average SFRs, contribute significantly to the total SFRD due to their large numbers and the presence of obscured star formation. Conversely, ELGs show a higher mean SFR per galaxy, particularly at low masses, placing them in the high sSFR regime.

Our results also reveal a mass-dependent shift in the utility of star formation tracers. The FIR emission becomes increasingly dominant and reliable with stellar mass due to enhanced dust attenuation. At the low-mass end, however, it fails to capture star formation activity, which is instead revealed by optical emission lines. As shown in Fig.~\ref{fig:sfr-per-bin}, including emission-line derived SFRs is necessary to trace this population and properly recover the low-mass end of the star-forming MS. To highlight this, we added a Popesso-style fit to the mean SFR per mass bin of the combined FIR and ELG sample, fixing $\log \mathrm{SFR}_{\max} = 1.51$ and $\gamma = 1$, in line with their formulation for the MS at $z = 0.9$. The agreement between our fit and the reference MS from \cite{popesso2023main}, along with the absence of significant deviations across the stellar mass range, suggests that the OTELO galaxies are consistent with other star-forming populations at a similar redshift and exhibit no notable evolutionary peculiarities.

Overall, this study demonstrates the need for a multi-wavelength approach to fully capture the complexity of star formation across galaxy populations. The FIR and optical tracers play complementary roles, with the former being most sensitive to massive dust-obscured galaxies and the latter being essential for probing star formation in low-mass, less obscured systems. Together, they provide a more complete and accurate view of star formation across cosmic time.

Looking forward, instruments such as the \textit{James Webb Space Telescope} \citep{gardner2012james, rigby2024jwst} and programs such as the Cosmic Evolution Early Release Science Survey \citep{finkelstein2023ceers} will greatly enhance IR-based studies of star formation, particularly by enabling the detection and characterisation of increasingly dust-obscured galaxies at NIR and MIR wavelengths. Together with FIR surveys such as the one presented in this work, these observations will provide a more complete view of obscured and unobscured star formation and will help refine our understanding of their interplay across cosmic time.

\begin{acknowledgements}

This research has been funded by grants PID2019\--107408GB-C41, PID-2021--122544NB--C41, PID--2021--122544NB--C43 and PID2022\--136598\-NB-C33 funded by MCIN/AEI/10.13039/501100011033 and by “ERDF A way of making Europe”.
Based on observations made with the Gran Telescopio Canarias (GTC), installed in the
Spanish Observatorio del Roque de los Muchachos of the Instituto de Astrof\'isica de
Canarias, in the island of La Palma.

This study makes use of data from AEGIS, a multi-wavelength sky survey conducted with the
\textit{Chandra}, GALEX, \textit{Hubble}, Keck, CFHT, MMT, Subaru, Palomar, \textit{Spitzer}, VLA, and other telescopes
and supported in part by the NSF, NASA, and the STFC.

Based  on  observations  obtained  with  MegaPrime/MegaCam,  a  joint  project  of  CFHT  and
CEA/IRFU, at the Canada-France-Hawaii Telescope (CFHT) which is operated by the National
Research Council (NRC) of Canada, the Institut National des Science de l'Univers of the
Centre National de la Recherche Scientifique (CNRS) of France, and the University of
Hawaii.  This work is based in part on data products produced at Terapix available at
the Canadian Astronomy Data Centre as part of the Canada-France-Hawaii Telescope Legacy
Survey, a collaborative project of NRC and CNRS.

Based on observations obtained with WIRCam, a joint project of CFHT,Taiwan, Korea, Canada,
France, at the Canada-France-Hawaii Telescope (CFHT) which is operated by the National
Research Council (NRC) of Canada, the Institute National des Sciences de l'Univers of the
Centre National de la Recherche Scientifique of France, and the University of Hawaii.
This work is based in part on data products produced at TERAPIX, the WIRDS (WIRcam Deep
Survey) consortium, and the Canadian Astronomy Data Centre. This research was supported by
a grant from the Agence Nationale de la Recherche ANR-07-BLAN-0228.

B.C. ~acknowledge the suppor of the Spanish Ministry of Science, Innovation and Universities through the project PID--2021--122544NB--C41.

J.N.~acknowledges the support of the National Science Centre, Poland through the SONATA BIS grant 2018/30/E/ST9/00208; Polish National Agency for Academic Exchange (NAWA) Bekker  grant BPN/BEK/2023/1/00271, and the kind hospitality of the IAC.

MGO acknowledges financial support from the State Agency for Research of the Spanish MCIU through Centre of Excellence Severo Ochoa award to the Instituto de Astrofísica de Andalucía CEX2021\-001131\-S funded by MCIN/AEI/10.13039/501100011033, and from the grant PID2022\- 136598NB\-C32 “Estallidos8”. MGO acknowledges support by the project ref. AST22-00001-Subp-11 funded from the EU\-NextGenerationEU.

MSP acknowledges the support of the Spanish Ministry of Science, Innovation and Universities through the project PID--2021--122544NB—C43.

JIGS acknowledges the support of the Spanish Ministry of Science, Innovation and Universities through the project PID--2021--122544NB—C44.

JGM ~acknowledge the suppor of the Spanish Ministry of Science, Innovation and Universities through the projects and PID--2021--122544NB--C41.

J.A.D acknowledges support from DGAPA-PAPIIT (UNAM), project IN116325.

\end{acknowledgements}

\bibliographystyle{aa}

\appendix

\section{Examples of SED fits}

To provide a clearer view of the fitting quality assessment described in Sect.~\ref{sec:SEDfitting}, 
we show here representative examples of sources meeting the selection criterion 
$\chi^2_{\rm red} < 2.5$ (“good” fits) and sources excluded from the sample due to poor fitting performance.

\subsection{Good fits: $\chi^2_{\rm red} < 2.5$}

\begin{figure}[h!]
\centering
\includegraphics[width=0.85\linewidth]{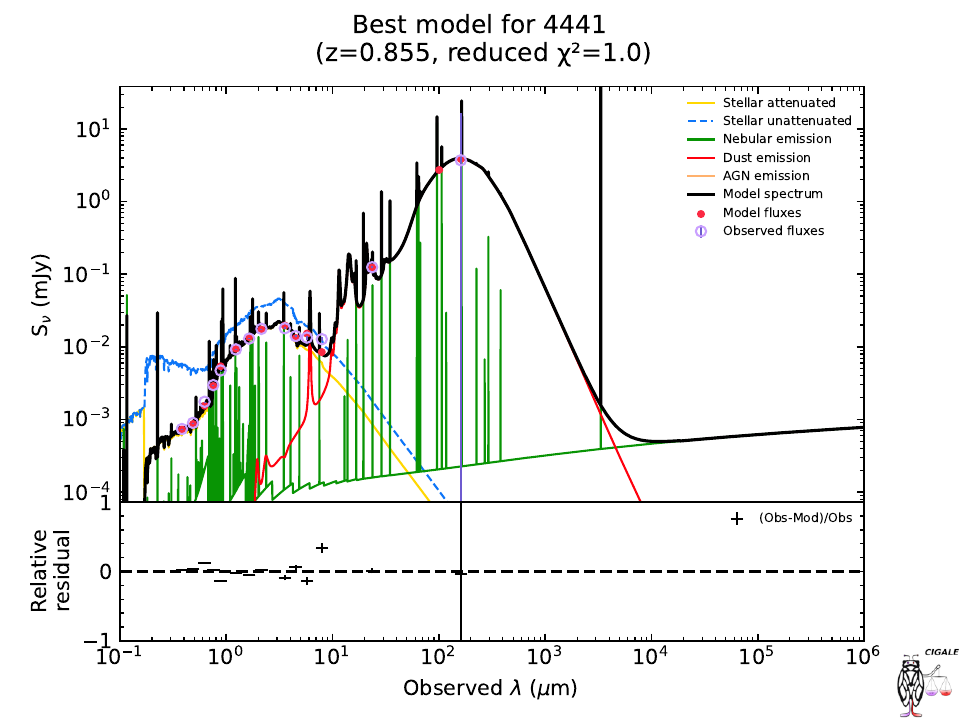}
\caption{Example of a good SED fit ($\chi^2_{\rm red} < 2.5$).}
\label{fig:goodfit1}
\end{figure}

\begin{figure}[h!]
\centering
\includegraphics[width=0.85\linewidth]{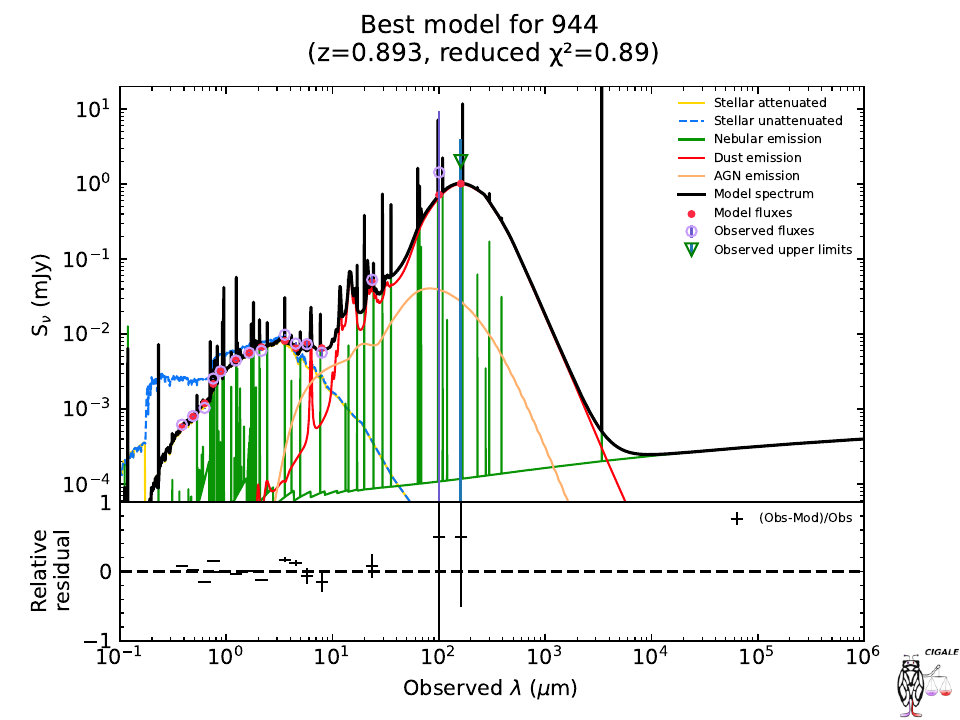}
\caption{Second example of a good SED fit.}
\label{fig:goodfit2}
\end{figure}

Figures~\ref{fig:goodfit1} and \ref{fig:goodfit2} show examples of SED fits that satisfy the 
$\chi^2_{\rm red} < 2.5$ requirement. These sources also visually follow the model predictions 
across the full wavelength range.

% Add more as needed:
% \includegraphics{goodfit3}, etc.

\subsection{Bad fits: Excluded sources}

\begin{figure}[h!]
\centering
\includegraphics[width=0.85\linewidth]{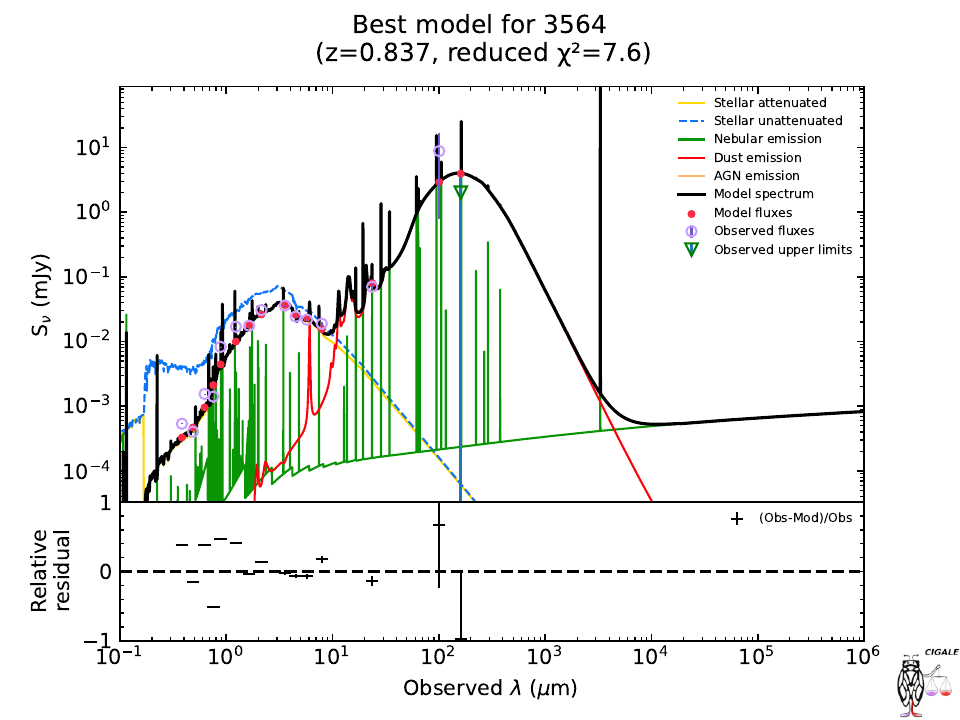}
\caption{Example of a rejected source due to poor fitting ($\chi^2_{\rm red} \sim 3$).}
\label{fig:badfit1}
\end{figure}

\begin{figure}[h!]
\centering
\includegraphics[width=0.85\linewidth]{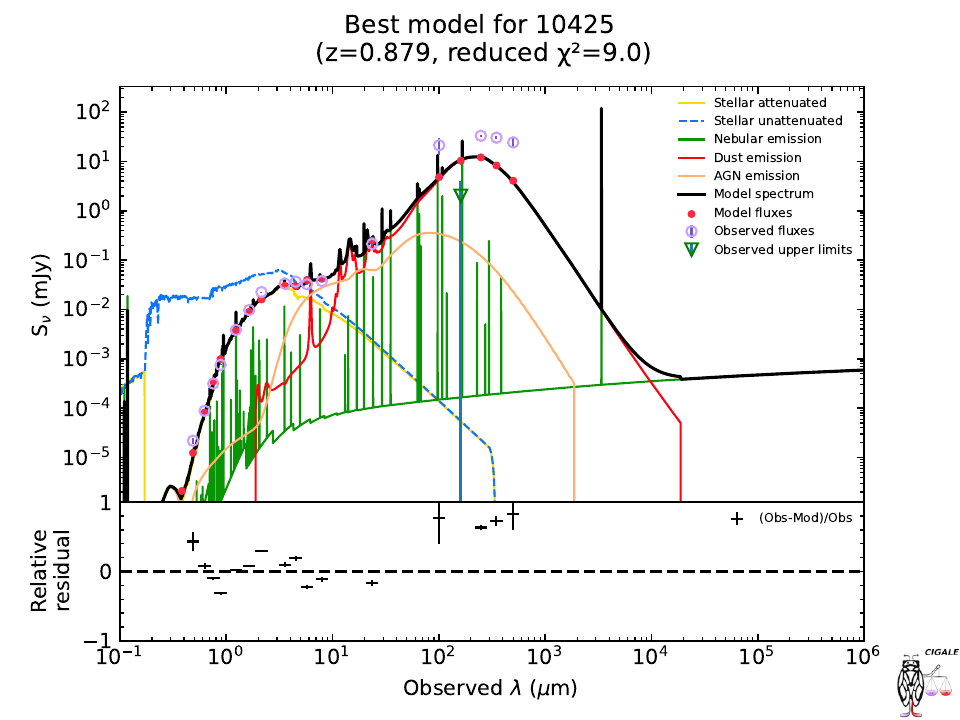}
\caption{Second example of a rejected source.}
\label{fig:badfit2}
\end{figure}

Figures~\ref{fig:badfit1} and \ref{fig:badfit2} present examples of excluded sources.  
These objects show either significantly higher reduced chi–square values or clear mismatches between the observed photometry and the model, justifying their removal from the final sample.

\end{document}